\documentclass[aps,prb,superscriptaddress,twocolumn,floatfix]{revtex4-1}

\usepackage{amsmath}
\usepackage{amssymb}
\usepackage{amsthm}
\usepackage{graphicx}
\usepackage{graphics}
\usepackage{subfigure}
\usepackage{color}
\usepackage{bm}
\usepackage{hyperref}
\usepackage{rotating}

\newcommand{\<}{\langle}
\renewcommand{\>}{\rangle}
\newcommand{\be}{\begin{equation} }
\newcommand{\ee}{\end{equation} }
\newcommand{\ba}{\begin{eqnarray} }
\newcommand{\ea}{\end{eqnarray} }

\newcommand{\bpm}{\begin{pmatrix}}
\newcommand{\epm}{\end{pmatrix}}
\newcommand{\bmm}{\begin{matrix}}
\newcommand{\emm}{\end{matrix}}

\def\t#1{\widetilde{#1}}
\def\wb#1{\overline{#1}}
\begin{document}

\title{Protected edge modes without symmetry}

\author{Michael Levin}
\affiliation{Condensed Matter Theory Center, Department of Physics, University of Maryland, College Park, 
Maryland 20742, USA}

%\date{\today}

\begin{abstract}
We discuss the question of when a gapped 2D electron system without any symmetry has a protected gapless edge
mode. While it is well known that systems with a nonzero thermal Hall conductance, $K_H \neq 0$, support
such modes, here we show that robust modes can also occur when $K_H = 0$ -- if the system has
quasiparticles with fractional statistics. We show that some types of fractional statistics
are compatible with a gapped edge, while others are fundamentally incompatible. More generally, we
give a criterion for when an electron system with abelian statistics and $K_H = 0$
can support a gapped edge: we show that a gapped edge is possible if and only if there
exists a subset of quasiparticle types $\mathcal{M}$ such that (1) all the quasiparticles in $\mathcal{M}$ have trivial
mutual statistics, and (2) every quasiparticle that is not in $\mathcal{M}$ has nontrivial mutual statistics
with at least one quasiparticle in $\mathcal{M}$. We derive this criterion using three different approaches:
a microscopic analysis of the edge, a general argument based on braiding statistics, and finally
a conformal field theory approach that uses constraints from modular invariance. We also discuss the
analogous result for 2D boson systems.
\end{abstract}

%\pacs{}

\maketitle

\section{Introduction} 
In two dimensions, some quantum many-body systems with a bulk energy gap have the property that they
support gapless edge modes which are extremely robust. These modes cannot be gapped out or localized by 
very general classes of interactions or disorder at the edge: they are ``protected'' by the structure of
the bulk phase. Examples include quantum Hall states,\cite{WenReview,WenBook} 
topological insulators,\cite{KaneMele,KaneMele2,HasanKaneRMP} and
topological superconductors,\cite{Schnyderetal} among others.

It is useful to distinguish between different levels 
of edge protection. In some systems, the edge excitations are only robust as long as certain symmetries are preserved. 
For example, in 2D topological insulators, the edge modes are protected by time reversal and charge conservation symmetry. 
If either of these symmetries are broken (either explicitly or spontaneously), the edge can be completely gapped. In 
contrast, in other systems, the edge modes are robust to arbitrary local interactions, independent of any symmetries.

While much previous work has focused on symmetry-protected edges, here we will focus on the latter, stronger, 
form of robustness. The goal of this paper is to answer a simple conceptual question: when does a gapped 2D 
quantum many-body system without any symmetry have a protected gapless edge mode?

One case in which such protected edge modes are known to occur is if the system has a nonzero thermal Hall 
conductance \cite{KaneFisherthermal,Kitaevhoneycomb} at low temperatures, i.e. $K_H \neq 0$. This result is 
particularly intuitive for systems whose edge can be modeled as a collection of chiral Luttinger liquids.
Indeed, in this case, 
$K_H = (n_L - n_R) \cdot \frac{\pi^2 k_B^2}{3h} T$, where $n_L, n_R$ are the number of left and right moving
chiral edge modes. Hence the condition $K_H \neq 0$ is equivalent to $n_L \neq n_R$. It is then clear
that $K_H \neq 0$ implies a protected edge: backscattering terms or other perturbations always gap out 
left and right moving modes in equal numbers, so if there is an imbalance between $n_L$ and $n_R$, the
edge can never be fully gapped. Alternatively, we can understand this result by analogy to 
the \emph{electric} Hall conductance, $\sigma_H$: just as systems with $\sigma_H \neq 0$ are guaranteed to have 
a gapless edge as long as charge conservation is not broken,\cite{Laughlinflux,Halperinedge} systems with 
$K_H \neq 0$ are guaranteed to have a gapless edge as long as energy conservation is not broken. 

On the other hand, if $K_H = 0$ then there isn't an obvious obstruction to gapping the edge. Thus, one might
guess that systems with $K_H = 0$ do not have protected edge modes. Indeed,
this is known to be true for systems of non-interacting fermions. \cite{Schnyderetal,Kitaevperiod} 

In this paper, we show that this intuition is incorrect in general: we find that systems with $K_H = 0$ 
can also have protected edge modes -- if they support quasiparticle excitations with \emph{fractional statistics}. 
The basic point is that some (but not all\cite{BravyiKitaev}) types of fractional statistics are fundamentally 
incompatible with a gapped edge. Thus, quasiparticle statistics provides another mechanism for edge protection which is 
qualitatively different from the more well-known mechanisms associated with electric or thermal Hall response.

Our main result is a criterion for when an electron system with abelian statistics and $K_H = 0$ 
can support a gapped edge (we discuss bosonic systems in the conclusion). We show that 
a gapped edge is possible if and only if there exists a set of quasiparticle ``types'' $\mathcal{M}$ satisfying two properties:
\begin{enumerate}
\item{
The particles in $\mathcal{M}$ have trivial mutual statistics: $e^{i\theta_{mm'}} = 1$ for any $m,m' \in \mathcal{M}$.
}
\item{
Any particle that is not in $\mathcal{M}$ has nontrivial mutual statistics with respect to at least one particle in $\mathcal{M}$: 
if $l \not\in \mathcal{M}$, then there exists $m \in \mathcal{M}$ with $e^{i\theta_{lm}} \neq 1$. 
}
\end{enumerate}
Here, two quasiparticle excitations are said to be of same topological ``type'' if they differ by an integer number of electrons. 
In this language, a gapped system typically has only a \emph{finite} set of distinct quasiparticle types, which we will 
denote by $\mathcal{L}$; the set $\mathcal{M}$ should be regarded as a subset of $\mathcal{L}$. Following previous terminology, 
\cite{Kapustintopbc} we will call any subset $\mathcal{M} \subseteq \mathcal{L}$ that obeys the above two properties a ``Lagrangian subgroup'' 
of $\mathcal{L}$.

Our analysis further shows that every gapped edge can be associated with a corresponding Lagrangian subgroup $\mathcal{M} \subseteq \mathcal{L}$. 
Physically, the set $\mathcal{M}$ describes the set of quasiparticles that can be ``annihilated'' at the edge, as explained in section
\ref{braidargsect}. We show that if $\mathcal{L}$ contains more than one Lagrangian subgroup, then the system supports more than one 
type of gapped edge: in general there is a different type of edge for every $\mathcal{M}$. In this sense, different types of 
gapped edges are (at least partially) classified by Lagrangian subgroups $\mathcal{M} \subseteq \mathcal{L}$.

We now briefly discuss the relationship with previous work on this topic. A systematic, microscopic analysis of gapped edges was 
presented in Ref. \onlinecite{KitaevKong}. In that work, the authors constructed and analyzed gapped edges for 
a large class of exactly soluble bosonic lattice models with both abelian and non-abelian quasiparticle statistics. On the other hand,
gapped edges were studied from a field theory perspective in Ref. \onlinecite{Kapustintopbc}. In that paper, the authors
investigated ``topological boundary conditions'' for abelian Chern-Simons theory.
Both analyses showed that gapped boundaries (or boundary conditions) are classified by an algebraic structure similar to the
Lagrangian subgroup $\mathcal{M} \subseteq \mathcal{L}$ introduced above. However, neither result implied that this classification scheme is general and includes
\emph{all} abelian gapped edges: indeed, it is not obvious, a priori, that exactly soluble models or topological boundary
conditions are capable of describing all types of gapped edges. One of the main contributions of this work is to fill in this hole 
and to show, in a concrete fashion, that every abelian gapped edge is associated with some Lagrangian subgroup $\mathcal{M} \subseteq \mathcal{L}$. 
It is this generality that allows us to deduce the existence of protected edges in those cases when $\mathcal{L}$ has no Lagrangian 
subgroup $\mathcal{M}$, i.e. when the criterion is violated.

We will derive the above criterion using three different approaches: a microscopic edge analysis, 
a general argument based on quasiparticle braiding statistics, and finally an 
argument that uses constraints from modular invariance. These derivations are complementary to one another.
The microscopic argument proves that the criterion is sufficient for having a gapped edge but does not prove
that it is necessary (it only provides evidence to that effect). The other two arguments show that the 
criterion is necessary for having a gapped edge, but do not prove that it is sufficient. 
%Also, while 
%the microscopic edge analysis gives a concrete picture of the protected edge modes, the braiding statistics 
%argument explains the physical meaning of the criterion, and the modular invariance derivation provides 
%another perspective based on conformal field theory.

This paper is organized as follows: in section \ref{examsect} we discuss some illustrative examples of the criterion, 
in sections \ref{microsect}-\ref{modinvsect} we establish the criterion with three different arguments, 
and in the conclusion we discuss the bosonic case and other generalizations. The appendix contains some of the more technical 
derivations.

\begin{figure}[tb]
\centerline{
\includegraphics[width=0.9\columnwidth]{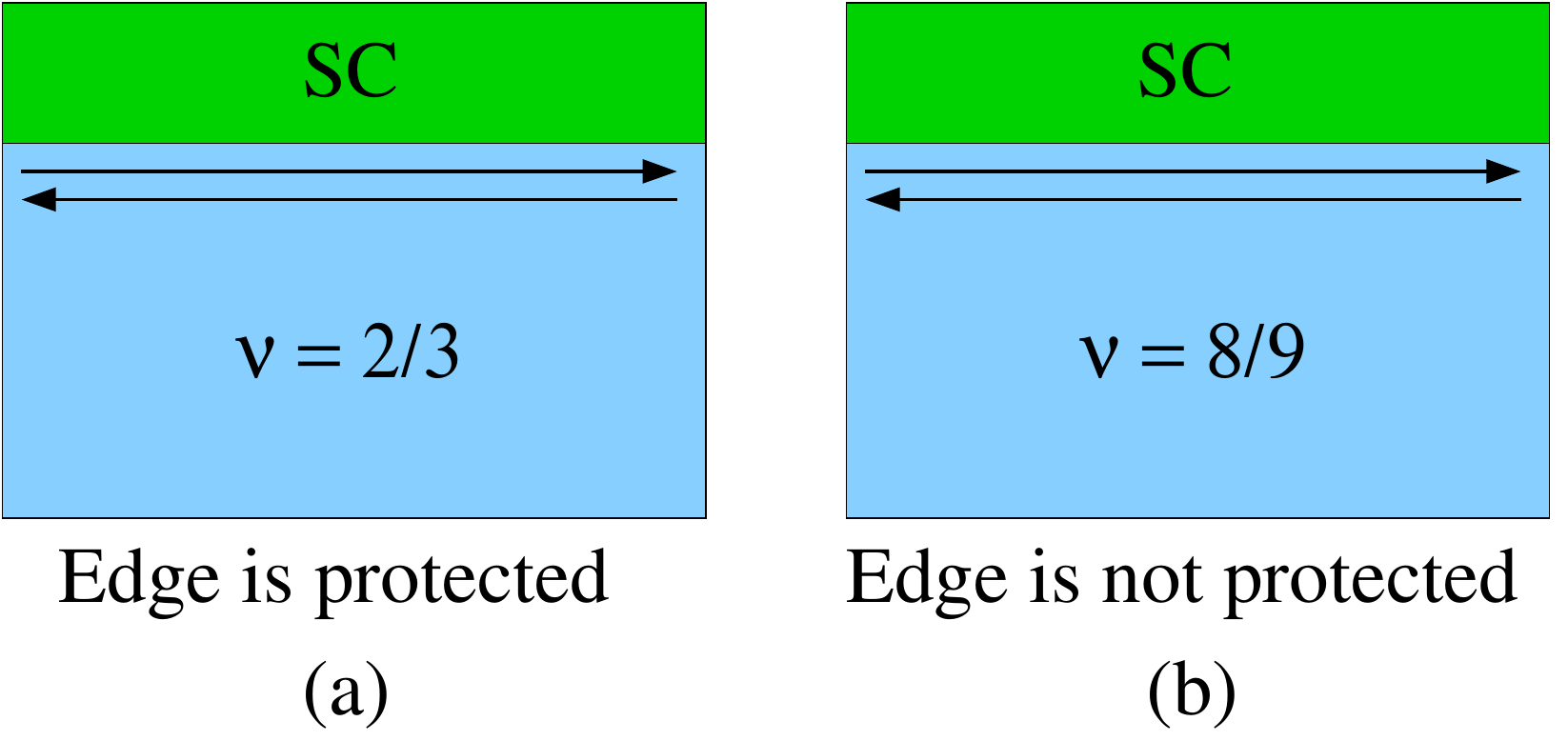}
}
\caption{
To demonstrate the subtleties of protected edge modes without symmetry, we consider the $\nu = 2/3$ and $\nu = 8/9$ fractional
quantum Hall edges, proximity-coupled to an adjacent superconductor.
(a) In the $\nu = 2/3$ case, the edge has vanishing thermal Hall conductance, $K_H = 0$, since it contains two 
modes moving in opposite directions. Even so, we will show that the edge is \emph{protected}. 
(b) In the $\nu = 8/9$ case, the edge also has $K_H = 0$, but in this case we will show that the edge is 
\emph{not protected}. We argue that the two states behave differently because of the different quasiparticle
braiding statistics in the bulk.
}
\label{qhefig}
\end{figure}

\section{Two examples} \label{examsect}
Before deriving the criterion, we first discuss a few examples that demonstrate its implications. 
A particularly illuminating example is the $\nu = 2/3$ fractional quantum Hall state -- that is, 
the particle-hole conjugate of the $\nu = 1/3$ Laughlin state. Let us consider a thought experiment in which 
the edge of the $\nu = 2/3$ state is proximity-coupled to a superconductor (Fig. \ref{qhefig}a). Then charge 
conservation is broken at the edge, so the edge does not have any symmetries. \footnote{We do not regard fermion parity 
conservation as a physical symmetry, as it cannot be broken by any local interactions.} At the same time, 
the thermal Hall conductance of this system vanishes since the edge has two modes that move in opposite directions. 
Thus, one might have guessed that the edge could be gapped by appropriate interactions. However, according
to the above criterion, this gapping is not possible: the edge is \emph{protected}. To see this, note that the $\nu = 2/3$ 
state supports $3$ different quasiparticle types which we denote by $\mathcal{L} = \{0, \frac{e}{3}, \frac{2e}{3} \}$. The 
mutual statistics of two quasiparticles $\frac{le}{3}, \frac{me}{3}$ is $\theta_{lm} = -\frac{2\pi l \cdot m}{3}$. 
Examining this formula, it is clear that $\mathcal{L}$ has no Lagrangian subgroup $\mathcal{M}$: the only set 
$\mathcal{M}$ that obeys condition (1) is $\mathcal{M} = \{0\}$ and this set clearly violates condition (2). 

To underscore the surprising nature of this result, it is illuminating to consider a second example: a 
$\nu = 8/9$ state constructed by taking the particle-hole conjugate of the $\nu = 1/9$ Laughlin state. (While
this state has not been observed experimentally, we can still imagine it as a matter of principle).  
Again, let us consider a setup in which the edge is proximity-coupled to a superconductor, thereby breaking charge
conservation symmetry (Fig. \ref{qhefig}b). This system is superficially very similar to the previous one, 
with two edge modes moving in opposite directions, and a vanishing thermal Hall conductance $K_H = 0$. However, 
in this case the above criterion predicts that the edge is \emph{not protected}. Indeed, the $\nu = 8/9$ state has 
$9$ different quasiparticle types which we denote by $\mathcal{L} = \{0,\frac{e}{9},\frac{2e}{9},...,\frac{8e}{9}\}$. The mutual 
statistics of two quasiparticles $\frac{le}{9}, \frac{me}{9}$ is given by $\theta_{lm} = -\frac{2\pi l \cdot m}{9}$. Examining this formula, 
we can see that the subset of quasiparticles $\mathcal{M} = \{0,\frac{3e}{9},\frac{6e}{9}\}$ obeys both (1) and (2), i.e. it is a valid Lagrangian
subgroup.

\section{Microscopic argument} \label{microsect}
In this section, we derive the criterion from a microscopic analysis of the edge. We 
follow an approach which is similar to that of Refs. \onlinecite{LevinStern,LevinSternlong, Neupertetal, 
XuMoore, LuVishwanath} and also the recent paper, Ref. \onlinecite{WangWen}. 

\subsection{Analysis of the examples} \label{examanalsect}
Before tackling the general case, we first warm up by studying the $\nu = 2/3$ and $\nu = 8/9$ examples discussed above.
Recall that the criterion predicts that the $\nu = 8/9$ edge can be gapped while the $\nu = 2/3$ edge is protected. We now
verify these claims by constructing edge theories for these two states and analyzing their stability.
  
We begin with the $\nu = 8/9$ state. To construct a consistent edge theory for this
state, consider a model in which there is a narrow strip of $\nu = 1$ separating the $\nu =8/9$ droplet 
and the surrounding vacuum. The edge then contains two chiral modes -- a forward propagating mode $\phi_1$ at 
the interface between the $\nu = 1$ strip and the vacuum, and a backward propagating mode $\phi_2$ at the 
interface between $\nu = 8/9$ and $\nu = 1$. The mode $\phi_1$ can be modeled as the usual 
$\nu =1$ edge: \cite{WenReview,WenBook}
\begin{equation}
L_1 = \frac{1}{4\pi} \left [ \partial_x \phi_1 \partial_t \phi_1 - v_1 (\partial_x \phi_1)^2 \right]
\end{equation}
Similarly, the mode $\phi_2$ can be modeled as the usual $\nu = 1/9$ edge, but with the opposite chirality:
\begin{equation}
L_2 = \frac{1}{4\pi} \left [ -9 \cdot \partial_x \phi_2 \partial_t \phi_2 - v_2 (\partial_x \phi_2)^2 \right ]
\end{equation}
Here, the two parameters $v_1, v_2$ encode the velocities of the two (counter-propagating) edge modes. We use a 
normalization convention where the electron creation operator corresponding to $\phi_1$ is
$\psi_1^\dagger = e^{i\phi_1}$, while the creation operator for $\phi_2$ is 
$\psi_2^\dagger = e^{-9i\phi_2}$. Combining these two edge modes into one Lagrangian $L = L_1 + L_2$ gives
\begin{equation}
L = \frac{1}{4\pi} \partial_x \phi_I (K_{IJ} \partial_t \phi_J - V_{IJ} \partial_x \phi_J)
\label{edgeth}
\end{equation}
where $I = 1,2$ and 
\begin{equation}
K = \bpm 1 & 0 \\ 0 & -9 \epm \ , \ V = \bpm v_1 & 0 \\ 0 & v_2 \epm
\label{K89}
\end{equation}
In this notation, a general product of electron creation and annihilation operators corresponds to an expression
of the form $e^{i\Lambda^T K \phi}$ where $\Lambda$ is a two component integer vector.

Given this setup, the question we would like to investigate is whether it is possible to gap out the above 
edge theory (\ref{edgeth}) by adding appropriate perturbations. For concreteness, we focus on perturbations
of the form 
\begin{equation}
U(\Lambda) = U(x) \cos(\Lambda^T K \phi - \alpha(x))
\label{tunnel}               
\end{equation}
where $\Lambda$ is a two component integer vector. These terms give an amplitude for electrons to
scatter from the forward propagating mode $\phi_1$ to the backward propagating mode $\phi_2$. Importantly,
we do not require $U(\Lambda)$ to conserve charge, since we are assuming charge conservation is broken by 
proximity coupling to a superconductor (Fig. \ref{qhefig}b). However, we do require that $U(\Lambda)$ conserve 
fermion parity.

We now consider the simplest scenario for gapping the edge: we imagine adding a \emph{single} backscattering term $U(\Lambda)$ 
to the edge theory (\ref{edgeth}). In this case, there is a simple condition that determines whether $U(\Lambda)$ can
open up a gap: according to the null vector criterion of Ref. \onlinecite{Haldanenull}, $U(\Lambda)$ 
can gap the edge if and only if $\Lambda$ satisfies
\begin{equation}
\Lambda^T K \Lambda = 0
\label{null}
\end{equation}
The origin of this criterion is that it guarantees that we can make a linear change of variables 
$\phi' = W \phi$ such that in the new variables, the edge theory (\ref{edgeth}) becomes a standard non-chiral
Luttinger liquid, and $U(\Lambda)$ becomes a backscattering term. It is then clear 
that the term $U(\Lambda)$ can gap out the edge, at least if $U$ is sufficiently large.\footnote{Note 
that it is not important whether $U(\Lambda)$ is relevant or irrelevant in the renormalization 
group sense. The reason is that we are not interested in the \emph{perturbative} stability of the 
edge, but rather whether it is stable to arbitrary local interactions.} Conversely, if 
$\Lambda$ doesn't satisfy (\ref{null}), it is not hard to show that the corresponding 
term $U(\Lambda)$ can \emph{never} gap out the edge, even for large $U$ (see appendix \ref{nullnecapp}).

Substituting (\ref{K89}) into (\ref{null}), and letting $\Lambda =(a,b)$ gives:
\begin{equation}
a^2 - 9 b^2 = 0
\end{equation}
By inspection, we easily obtain the solution
$\Lambda = (3,-1)$. It follows that the corresponding scattering term $U(\Lambda)$ can gap out the edge. 
We note that this term is not charge conserving, since it corresponds to a process in which one electron is annihilated
on one edge mode and three are created on the other. However, it is still an allowed perturbation in the presence of
the superconductor, since it conserves fermion parity (in fact, it is not hard to show that solutions to 
(\ref{null}) always conserve fermion parity).

Now let us consider the $\nu = 2/3$ edge. Following a construction similar to the one outlined above, we model 
the edge by the theory (\ref{edgeth}) with
\begin{equation}
K = \bpm 1 & 0 \\ 0 & -3 \epm
\label{K23}
\end{equation}
As before, we ask whether backscattering terms $U(\Lambda)$ can gap the edge, and as 
before, we can answer this question by checking whether $\Lambda$ satisfies the null vector condition 
(\ref{null}). However, in this case, we can see that (\ref{null}) reduces to
\begin{equation}
a^2 - 3 b^2 = 0
\end{equation}
which has \emph{no integer solutions}, since $\sqrt{3}$ is
irrational. We conclude that no single perturbation $U(\Lambda)$ can open up a gap -- suggesting that the edge is protected.

We emphasize that this analysis only shows that the $\nu = 2/3$ edge is robust against a particular class of 
perturbations -- namely single backscattering terms of the form (\ref{tunnel}). Hence, the above derivation only
gives \emph{evidence} that $\nu = 2/3$ is protected; it does not prove it. 

\subsection{General abelian states} \label{gensect}
We now extend the above analysis to general electron systems with abelian quasiparticle statistics
and with $K_H = 0$. For each state, we investigate whether its edge modes can be gapped out by 
simple perturbations. We show that if a state satisfies the criterion,
then its edge can be gapped out. Conversely, we show that if a state does not satisfy the criterion then its edge 
is protected -- at least against the perturbations considered here. In this way, we prove that the criterion
is \emph{sufficient} for having a gapped edge, and we give evidence that it is necessary.

Our analysis is based on the Chern-Simons framework for describing gapped abelian states of matter. According to this framework, 
every abelian state can be described by a $p$ component $U(1)$ Chern-Simons theory of the form\cite{WenBook,WenReview,WenKmatrix}
\begin{equation}
L_{B} = \frac{K_{IJ}}{4\pi} \epsilon^{\lambda \mu \nu} a_{I \lambda} \partial_\mu a_{J \nu}
\label{cstheory}
\end{equation}
where $K$ is a symmetric, non-degenerate $p \times p$ integer matrix. In this formalism, the quasiparticle excitations
are described by coupling $L_B$ to bosonic particles that carry integer gauge charge $l_I$ under each of the gauge fields
$a_I$. Thus, the quasiparticle excitations are parameterized by $p$ component integer vectors $l$. 
The mutual statistics of two excitations $l,l'$ is given by
\begin{equation}
\theta_{ll'} = 2\pi l^T K^{-1} l'
\end{equation}
while the exchange statistics is $\theta_l = \theta_{ll}/2$. Excitations 
composed out of electrons correspond to vectors $l$ of the form 
$l = K \Lambda$ where $\Lambda$ is a $p$ component integer vector. 
Two quasiparticle excitations $l,l'$ are ``equivalent'' or ``of the same type'' 
if they differ by some number of electrons, i.e. $l - l' = K \Lambda$ for some $\Lambda$. 

In this paper, since we are interested in states with equal numbers of left 
and right moving edge modes (i.e. states with $K_H = 0$), we will restrict ourselves to $K$-matrices with 
vanishing signature and dimension $p=2N$. \footnote{It can be shown that the signature of $K$ is equal 
to the difference between the number of left and right moving edge modes, $n_L - n_R$, using the bulk-edge correspondence described below.} 
Also, since we wish to study states built out of electrons, we focus on $K$-matrices with at least one odd element
on the diagonal: this assumption guarantees that the Chern-Simons theory (\ref{cstheory}) supports at 
least one topologically trivial excitation with fermionic statistics -- i.e. at least one
electron-like excitation. (Likewise, when we study bosonic states in appendix \ref{boseapp}, we 
consider $K$-matrices with only \emph{even} elements on the diagonal).

Let us translate the criterion from the introduction into the $K$-matrix language. The set of 
quasiparticles $\mathcal{M}$ corresponds to a collection of (inequivalent) $2N$ component integer vectors, 
$\mathcal{M} = \{m\}$. Conditions (1) and (2) translate to the requirements that:
\begin{enumerate}
\item{$m^T K^{-1} m'$ is an integer for any $m, m' \in \mathcal{M}$.}
\item{If $l$ is not equivalent to any element of $\mathcal{M}$, then $m^T K^{-1} l$ is 
non-integer for some $m \in \mathcal{M}$.}
\end{enumerate}
The criterion states that the edge can be gapped if and only if there exists a set 
$\mathcal{M}$ satisfying these two conditions.

To derive this result, we use the bulk-edge correspondence for abelian Chern-Simons theory \cite{WenReview,WenBook} 
to model the edge as a $2N$ component chiral boson theory (\ref{edgeth}). We then ask whether the edge can be
gapped by adding backscattering terms $U(\Lambda)$ (\ref{tunnel}). In order to gap out all $2N$ edge 
modes, we need $N$ terms, $\sum_{i=1}^N U(\Lambda_i)$, where $\{\Lambda_1,...,\Lambda_N\}$ are all linearly independent. 
Similarly to Eq. \ref{null}, there is a simple condition for when the perturbation $\sum_{i=1}^N U(\Lambda_i)$ 
can gap out the edge. Specifically, one can show that this term can gap out the edge
if and only if $\{\Lambda_1, ..., \Lambda_N\}$ satisfy
\begin{equation}
\Lambda_i^T K \Lambda_j = 0
\label{gennull}
\end{equation}
for all $i,j$. 

To complete the derivation, we make use of a mathematical result derived in appendix \ref{equivapp}.
According to this result, equation (\ref{gennull}) has a solution $\{\Lambda_1,...,\Lambda_N\}$ if and only if there exists 
a set of integer vectors $\mathcal{M}$ with the above two properties. (More generally, appendix \ref{equivapp}
establishes a correspondence between sets of null vectors $\{\Lambda_i\}$ and Lagrangian subgroups $\mathcal{M}$).

Putting this all together, we arrive at two conclusions. First, every state that satisfies 
the criterion can support a gapped edge. Second, every state that does not satisfy criterion 
has a protected edge -- at least with respect to perturbations of the form $\sum_{i=1}^N U(\Lambda_i)$.

\section{Braiding statistics argument} \label{braidargsect}
The above microscopic derivation leaves several questions unanswered. First, it does not explain the
\emph{physical} connection between bulk braiding statistics and protected edge modes. Instead, this 
connection emerges from a mathematical relationship between null vectors $\{\Lambda_i\}$
and Lagrangian subgroups $\mathcal{M}$. Another problem is that the derivation is 
not complete since it only analyzes the robustness of the edge with respect to a particular class of 
perturbations. As a result, we have not proven definitively that the criterion is \emph{necessary} for
having a gapped edge. In this section, we address both of these problems: we give a general 
argument showing that any system that supports a gapped edge \emph{must} satisfy the criterion. In addition,
this argument reveals the physical meaning of the set $\mathcal{M}$.  

\begin{figure}[tb]
\centerline{
\includegraphics[width=0.9\columnwidth]{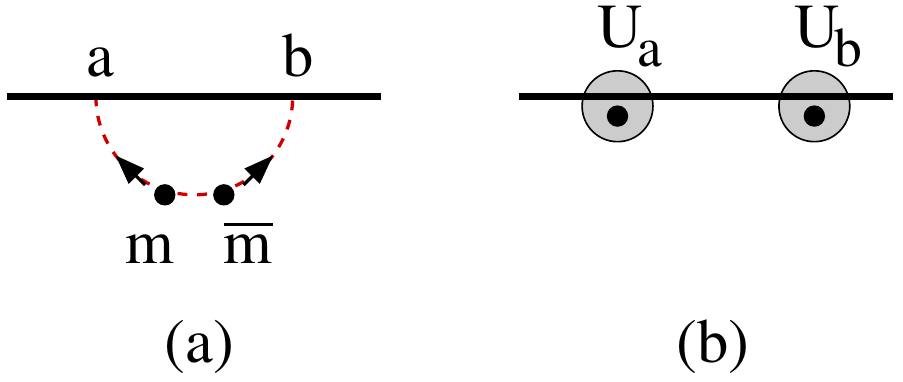}
}
\caption{
The concept of ``annihilating'' particles at a gapped boundary.
(a) Consider a thought experiment in which we create a pair of quasiparticle excitations $m, \wb{m}$ in the bulk 
and then bring them near to two points $a,b$ at the edge. We denote the resulting excited state by $|\Psi_{ex}\>$.
(b) We say that $m, \wb{m}$ can be annihilated at the boundary if there exist operators $U_a, U_b$
acting in the vicinity of $a,b$, such that $U_a U_b |\Psi_{ex}\> = |\Psi\>$, where $|\Psi\>$ is the ground state.
Otherwise we say the particles cannot be annihilated.
}
\label{braidfig}
\end{figure}

We begin by explaining the notion of ``annihilating'' quasiparticles at a gapped boundary. The idea is as 
follows. Consider a general gapped electron system with a gapped boundary. Let us 
imagine that we take the ground state $|\Psi\>$ and then excite the system by creating a quasiparticle/quasihole pair 
$m, \wb{m}$ somewhere in the bulk. After creating these excitations, we separate them and then bring them near 
two points $a,b$ on the edge (Fig. \ref{braidfig}a). Let us denote the resulting state by $|\Psi_{ex}\>$. We will 
say that $m, \wb{m}$ ``can be annihilated at the boundary'' if, for arbitrarily distant $a,b$, there 
exist operators $U_a, U_b$ acting in finite regions near $a,b$, such that (Fig. \ref{braidfig}b)
\begin{equation}
U_a U_b |\Psi_{ex}\> = |\Psi\>
\end{equation} 
Likewise, if no such operators exist then we will say 
that $m, \wb{m}$ cannot be annihilated at the boundary. Here, $U_a$ and $U_b$ can be any operators composed out 
electron creation and annihilation operators acting in the vicinity of $a$ and $b$ such that 
$U_a \cdot U_b$ conserves fermion parity. We note that we do not require that $U_a$ and $U_b$ individually conserve 
fermion parity -- only that their product $U_a \cdot U_b$ does so. Thus, according to the above definition, 
electron-like excitations can always be annihilated at the boundary, e.g. via 
$U_a = c_a$, $U_b = c_b^\dagger$.

Let $\mathcal{M}$ be the set of all quasiparticle types that can be annihilated at the edge: 
\begin{equation}
\mathcal{M} = \{m : m \text{ can be annihilated at edge}\}
\end{equation}
We will now argue that self-consistency requires that $\mathcal{M}$ has a very special structure: in particular, for 
systems with abelian quasiparticle statistics, $\mathcal{M}$ must be a Lagrangian subgroup.
In other words, we will show that (1) any two quasiparticle types that can be annihilated at the
edge must have trivial mutual statistics, and (2) any quasiparticle type that cannot be annihilated must have
nontrivial statistics with at least one particle that can be annihilated. This will establish that the 
criterion is \emph{necessary} for having a gapped edge.

\begin{figure}[tb]
\centerline{
\includegraphics[width=0.9\columnwidth]{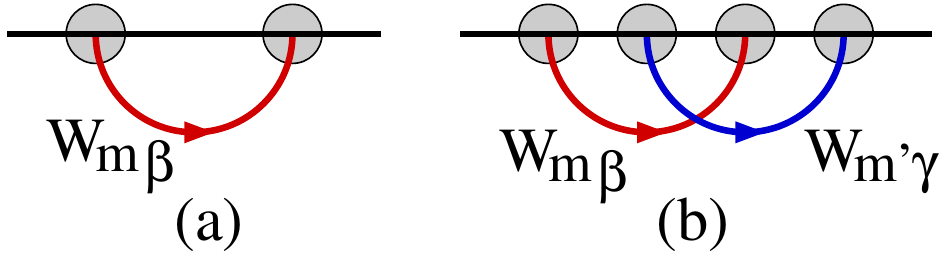}
}
\caption{
(a) For each $m \in \mathcal{M}$, we can consider a process in which we create a pair of quasiparticles 
$m,\wb{m}$ in the bulk, move them along some path $\beta$ to two points on the edge, and then annihilate them. 
We define $\mathbb{W}_{m\beta}$ to be the operator that implements this process.
(b) To establish condition (1) of the criterion, we consider two paths $\beta, \gamma$, and two associated operators
$\mathbb{W}_{m\beta}$, $\mathbb{W}_{m'\gamma}$. We then make use of the relations (\ref{wmm1}-\ref{wmm2}) along with 
the commutation relation (\ref{statrel}). 
}
\label{wmfig}
\end{figure}

We establish condition (1) using an argument similar to one given in Ref. \onlinecite{LevinGu}. The first step
is to consider a three step process in which we create two 
quasiparticles $m,\wb{m}$ in the bulk, move them along some path $\beta$ to two points on the edge, and then 
annihilate them. At a formal level, this process can be implemented by multiplying the
ground state $|\Psi\>$ by an operator of the form
\begin{equation}
\mathbb{W}_{m \beta} = U_a U_b W_{m \beta}
\label{wmbetadef}
\end{equation} 
Here, $W_{m \beta}$ is a (string-like) unitary operator that creates the quasiparticles and moves them to the edge, 
while $U_a U_b$ is an operator that annihilates them (Fig. \ref{wmfig}a). Given that the system returns to 
the ground state at the end of the process, we have the algebraic relation
\begin{equation}
\mathbb{W}_{m \beta} |\Psi\> = |\Psi\>
\label{onew}
\end{equation}
(Here we assume that the phase of the operator $\mathbb{W}_{m \beta}$ has been adjusted so that there is no phase factor on the 
right hand side of Eq. \ref{onew}).

Now imagine we repeat this process for another quasiparticle $m'$ and another path $\gamma$ with endpoints $c,d$
(Fig. \ref{wmfig}b). We denote the corresponding operator by 
\begin{equation}
\mathbb{W}_{m' \gamma} = U_c U_d W_{m' \gamma}
\end{equation}
Since each 
process returns the system to the ground state, we have: 
\begin{equation}
\mathbb{W}_{m'\gamma} \mathbb{W}_{m\beta} |\Psi\> = |\Psi\>
\label{wmm1}
\end{equation}
Similarly, if we execute the processes in the opposite order, we have
\begin{equation}
\mathbb{W}_{m\beta} \mathbb{W}_{m'\gamma} |\Psi\> = |\Psi\>
\label{wmm2}
\end{equation}
At the same time, it is not hard to see that
$\mathbb{W}_{m\beta}, \mathbb{W}_{m'\gamma}$ satisfy the commutation algebra
\begin{equation}
\mathbb{W}_{m\beta} \mathbb{W}_{m'\gamma} |\Psi\> = e^{i\theta_{mm'}} 
\mathbb{W}_{m'\gamma} \mathbb{W}_{m\beta} |\Psi\>
\label{statrel}
\end{equation}
where $e^{i\theta_{mm'}}$ is the mutual statistics between $m,m'$. Indeed, it
follows from a general analysis of abelian quasiparticle statistics that
\begin{equation}
W_{m\beta} W_{m'\gamma} |\Psi\> = e^{i\theta_{mm'}} W_{m'\gamma} W_{m\beta} |\Psi\>
\label{stringcomm}
\end{equation}
for any two paths $\beta, \gamma$ that intersect one another at one point
(see e.g. Refs. \onlinecite{LevinGu,LevinWenHop}). Using this result, together with the
observation that $W_{m'\gamma}$ commutes with $U_a U_b$ and $W_{m \beta}$ commutes with
$U_c U_d$ (since they act on non-overlapping regions), equation (\ref{statrel}) follows immediately.

In the final step, we compare (\ref{statrel}) with (\ref{wmm1}-\ref{wmm2}). Clearly,
consistency requires that $e^{i\theta_{mm'}} = 1$ for all $m,m' \in \mathcal{M}$. Hence $\mathcal{M}$ must satisfy condition (1).

\begin{figure}[tb]
\centerline{
\includegraphics[width=0.9\columnwidth]{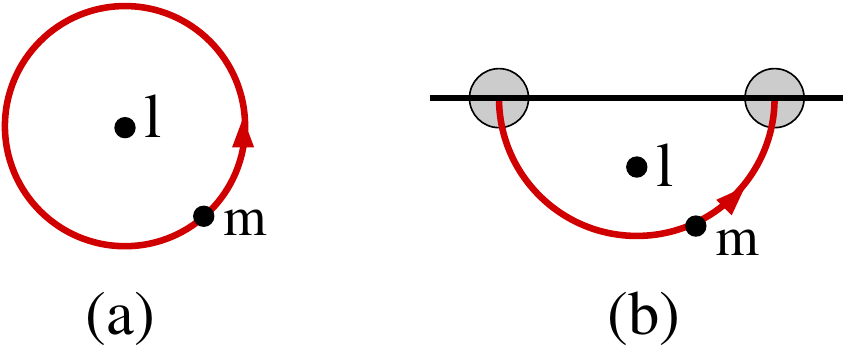}
}
\caption{
(a) The concept of braiding non-degeneracy in the \emph{bulk}: if $l$ is a quasiparticle that cannot be
annihilated in the bulk, then there must be at least one quasiparticle species $m$ that has nontrivial mutual
statistics with respect to $l$, i.e. $e^{i\theta_{lm}} \neq 1$.
(b) The concept of braiding non-degeneracy at a gapped \emph{edge}: if $l$ cannot be annihilated at the edge, then there
must be at least one quasiparticle $m$ that can be annihilated at the edge such that $e^{i\theta_{lm}} \neq 1$.
}
\label{brnondegdeffig}
\end{figure}

Showing that $\mathcal{M}$ satisfies condition (2) is more challenging. Here we simply explain the intuition behind
this claim; in appendix \ref{braidnondegapp} we give a detailed argument. To begin, we recall a \emph{bulk} property of systems with
fractional statistics known as ``braiding non-degeneracy'' (appendix E.5 of Ref. \onlinecite{Kitaevhoneycomb}). Suppose 
$l$ is a quasiparticle excitation that cannot be annihilated \emph{in the bulk}. That is, suppose that if we create 
$l, \wb{l}$ out of the ground state and the bring them near two widely separated points $a,b$ in the bulk, then we 
cannot annihilate them by applying appropriate operators $U_a, U_b$ acting in their vicinity. Braiding non-degeneracy 
is the statement that, for any such $l$, there is always at least one quasiparticle $m$ that has nontrivial mutual 
statistics with respect to $l$, i.e., $e^{i\theta_{lm}} \neq 1$ (Fig. \ref{brnondegdeffig}a). 

The intuition behind braiding non-degeneracy is as follows: if $l$ cannot be annihilated by applying any operator, then
in particular it cannot be annihilated by cutting a large hole around $l$. Hence it must be possible to detect the 
presence of this excitation outside any finite disk centered at $l$. At the same time, it is natural to expect that 
the only way to detect excitations non-locally is by an Aharonov-Bohm measurement -- i.e. braiding quasiparticles around them 
and measuring the associated Berry phase. Putting these two observations together, we deduce that 
$e^{i\theta_{lm}} \neq 1$ for some $m$ since otherwise it would not be possible to detect $l$ in this way.

For the same reason that bulk fractionalized systems obey braiding non-degeneracy, it is natural to expect that
the gapped \emph{edges} of these systems should obey an analogous property. Specifically, we expect that
\emph{for each quasiparticle $l$ which cannot be annihilated at the edge, there
must be at least one quasiparticle species $m$ which can be annihilated at the edge and which satisfies 
$e^{i\theta_{lm}} \neq 1$}. The physical intuition behind this statement is similar to that of bulk braiding 
non-degeneracy: we note that if $l$ cannot be annihilated, it must be detectable 
by a measurement far from $l$. Again, it is reasonable to expect that this non-local detection is based on 
braiding, since both the edge and bulk are gapped and hence\cite{Hastingscorr} have a finite correlation 
length. In an edge geometry, the analogue of conventional braiding is 
to create a pair of quasiparticles $m, \wb{m}$ in the bulk, bring them to the edge on either side of $l$ and annihilate them 
(Fig. \ref{brnondegdeffig}b). Assuming that it is possible to detect $l$ in this way, it follows that there 
always exists at least one quasiparticle $m$ which can be annihilated at the edge, and which has 
$e^{i\theta_{lm}} \neq 1$. (See appendix \ref{braidnondegapp} for a detailed argument). 
This result is exactly the statement that $\mathcal{M}$ satisfies condition (2). We conclude that $\mathcal{M}$ is a Lagrangian 
subgroup, as claimed.

The reader may wonder: at what point in the argument do we use the assumption that the edge is gapped? This
assumption enters in several ways. At an intuitive level, it is implicit in the very definition of quasiparticle
annihilation: the physical picture of annihilating quasiparticles with (exponentially) localized operators $U_a, U_b$ 
is only sensible if the edge has a finite correlation length. If instead the correlation 
length were infinite -- as is typical for a gapless edge -- then we would not expect such a 
localized annihilation process to be possible in general. At a mathematical level, the gapped edge assumption 
plays an important role in the derivation of condition (2): only for a gapped edge can one establish an analogue of braiding 
non-degeneracy (see appendix \ref{braidnondegapp}).

\section{Modular invariance argument} \label{modinvsect}
To complete our discussion, we present another argument that shows that the criterion is necessary for gapping
the edge. In order to understand this argument, it is helpful to consider it in a larger context. Recall that there 
is a close relationship between protected edge modes in $2D$ systems and ``no-go'' theorems about $1D$ 
lattice models. For each type of protected edge, there is typically a corresponding no-go
theorem ruling out the possibility of constructing a $1D$ lattice model realizing that edge theory. 
For example, corresponding to the integer quantum Hall edge is a theorem\cite{Friedan} that states
that it is impossible to construct a $1D$ lattice model realizing a chiral fermion.

In some cases it is possible to use a $1D$ no-go theorem to \emph{prove} that a $2D$ edge is protected.
This is the strategy we will follow here. The no-go theorem we use is the statement that \emph{any conformal field theory 
(CFT) realized by an $1D$ lattice model must be modular invariant} -- i.e. it is impossible to realize a CFT that
violates modular invariance in a $1D$ system. \cite{Francesco} Using this theorem (or more accurately, 
conjecture), we prove that the criterion is necessary for having a gapped edge. We note that this
modular invariance approach is similar to that of Ref. \onlinecite{RyuZhang}. (See also Refs. 
\onlinecite{CappelliZemba, CappelliViolaZemba,CappelliViola} for related work).

We proceed in the same way as in the previous section: we consider a general gapped electron system
that has abelian quasiparticle statistics, has $K_H = 0$, and supports a gapped edge. We then 
show that the set $\mathcal{M}$ of particles that can be annihilated
at the edge must be a Lagrangian subgroup, i.e. must obey conditions (1) and (2) of the criterion. 
 
\begin{figure}[tb]
\centerline{
\includegraphics[width=0.95\columnwidth]{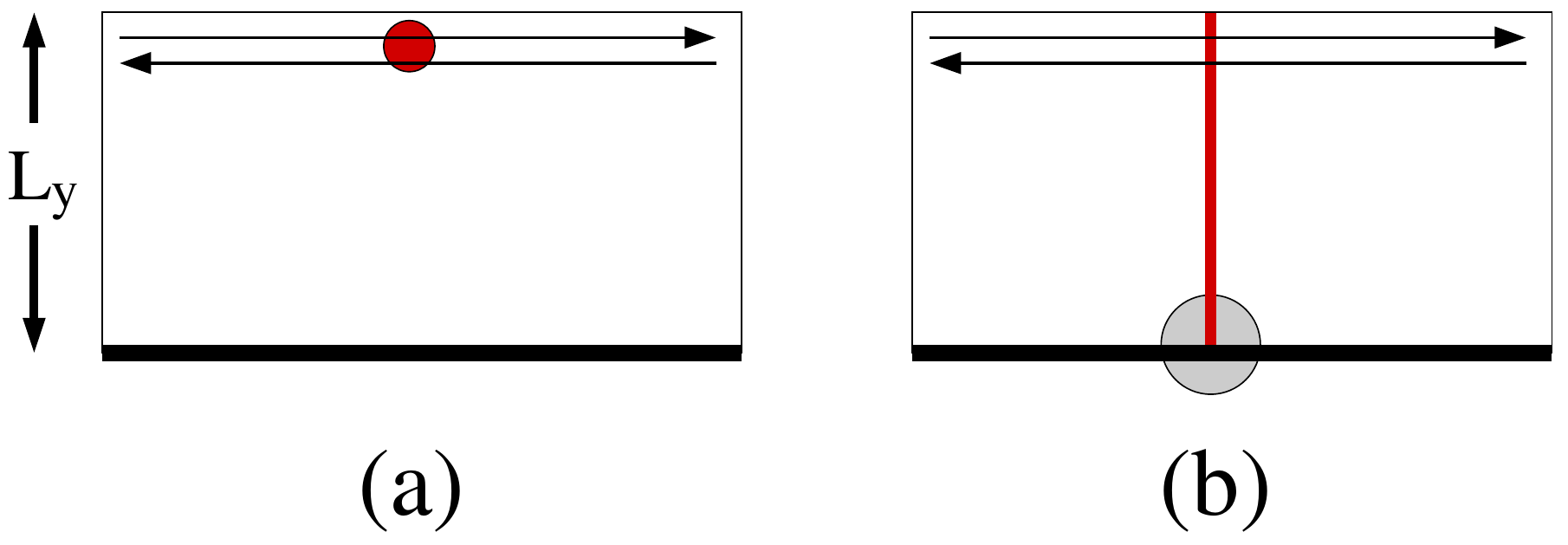}
}
\caption{
We consider the system in a strip geometry with finite width $L_y$ in the $y$-direction.
We assume that the lower edge is gapped, while the upper edge is gapless and described by
(\ref{genedgeth}).
The system has two types of scaling operators: (a) charge neutral operators (\ref{scalop1}) acting on
the upper edge and (b) charged operators of the form $e^{il^T \phi}$. Operators of type (b) can only appear
as a low-energy description of a tunneling process in which a quasiparticle of type $l$ tunnels from the upper edge
to the lower edge and is subsequently annihilated.
}
\label{cftoperfig}
\end{figure}

The starting point for the argument is to consider the system in a strip geometry with a large but finite width $L_y$ 
in the $y$ direction and infinite extent in the $x$ direction. Since the system supports a gapped edge, we can consider
a scenario in which the lower edge is gapped. At the same time, we imagine tuning the interactions at the upper edge so that
it is gapless (Fig. \ref{cftoperfig}). Specifically, we tune the interactions so that the upper edge is described by the model edge theory 
\begin{equation}
L = \frac{1}{4\pi} \partial_x \phi_I (K_{IJ} \partial_t \phi_J - V_{IJ} \partial_x \phi_J)
\label{genedgeth}
\end{equation}
where $K_{IJ}$ is the $2N \times 2N$ $K$-matrix describing the bulk system.

To proceed further, we make a change of variables to diagonalize the above action. Let $W$ be a real matrix such that 
$W^T K W= \Sigma_z$ where $\Sigma_z = \bpm \bf{1} & 0 \\ 0 & -\bf{1} \epm$ and $\bf{1}$ denotes the 
$N \times N$ identity matrix. Setting $\phi_I =  W_{IJ} \tilde{\phi}_J$, the edge theory becomes
\begin{equation}
L = \frac{1}{4\pi} \partial_x \t{\phi}_I (\Sigma_z \partial_t \t{\phi}_J - \t{V}_{IJ} \partial_x \t{\phi}_J)
\label{cftedge}
\end{equation}
where $\t{V} = W^T V W$. If we tune the interactions at the upper edge appropriately, we can arrange so that 
$\t{V}$ is of
the form $\t{V} = v \delta_{IJ}$. Then all the edge modes propagate at the same speed $|v|$ and the low energy, 
long wavelength physics of the strip is described by a \emph{conformal field theory}. 

We now apply the no-go theorem discussed above: we note that the strip is a quasi-$1D$ system, so according to the 
no-go theorem/conjecture, the above conformal field theory must be modular invariant. Our basic strategy
will be to use this modular invariance constraint to derive the criterion. 

Before doing this, we first briefly review the definition of modular invariance (for a more detailed discussion 
see e.g. Ref. \onlinecite{Francesco}). For any conformal field theory, we can imagine listing all the scaling operators 
$\mathcal{O}$ along with their scaling dimensions $(\Delta, \wb{\Delta})$ defined by 
\begin{equation}
\<\mathcal{O}(0,0)\mathcal{O}(x,t)\> \sim \frac{1}{(x-vt)^{2\Delta}} \cdot \frac{1}{(x+vt)^{2\wb{\Delta}}}
\end{equation} 
Using this list, we can construct the formal sum (``partition function'')
\begin{equation}
Z(\tau) = e^{\pi i c(\wb{\tau} - \tau)/12} \sum_{\mathcal{O}} e^{2\pi i (\Delta \tau - \wb{\Delta} \wb{\tau})}
\label{partfunct}
\end{equation}
where $c$ is the central charge and $\tau$ is a formal parameter. If we evaluate this expression for a 
complex $\tau$ with $Im(\tau) > 0$, the sum converges.  Modular invariance is the statement 
that $Z(\tau)$ has to obey the constraint 
\begin{equation}
Z(-1/\tau) = Z(\tau) \label{modinv} 
\end{equation}
This equation places restrictions on the operator content of the conformal field theory -- that is, the 
set of scaling operators in the theory. To see where it comes from, we note that $Z(\tau)$ can be interpreted
physically as the euclidean space-time partition function for the conformal field theory, evaluated on a torus 
of shape $\tau$. Equation \ref{modinv} then follows from the fact that the two toruses with shape $\tau$ and $-1/\tau$ are 
conformally equivalent to one another and therefore must give identical partition functions. (For similar
reasons, modular invariance also imposes the constraint that $Z(\tau+1) = Z(\tau)$ for a bosonic system and 
$Z(\tau+2) = Z(\tau)$ for a fermionic system, but we will not need this result here).

We now investigate the implications of modular invariance, in particular equation (\ref{modinv}), for our system.
The first step is to classify all the scaling operators and find their scaling dimensions. Importantly, we should only 
consider scaling operators which are \emph{local} in the $x$-direction -- that is, operators composed out of products 
of electron creation and annihilation operators acting within some finite segment of the strip 
$[x-\Delta x,x+\Delta x]$.

One set of scaling operators is given by expressions of the form
\begin{equation}
\mathcal{O}_{\{n_{J,k}\}} = \prod_{J=1}^{2N} \prod_{k=1}^\infty (\partial^k_x \t{\phi}_J)^{n_{J,k}}
\label{scalop1}
\end{equation}
These operators describe combinations of electron creation and annihilation operators which are charge neutral
in each individual edge mode $\t{\phi}_J$ (Fig. \ref{cftoperfig}a). Another set of operators are expressions of 
the form $e^{i l^T \phi}$ for integer vectors $l$. These operators describe the annihilation (or creation) 
of a quasiparticle of type $l$ on the upper edge. An important point is that not all $l$ correspond to physical 
operators. Indeed, in general one cannot annihilate a fractionalized quasiparticle by itself. The only way such 
an operator can appear in our theory is as a description of a tunneling/annihilation process in which a 
quasiparticle of type $l$ tunnels from the upper edge to the lower edge and is subsequently annihilated 
(Fig. \ref{cftoperfig}b). Thus, \emph{the allowed values of $l$ correspond to the quasiparticles that can be 
annihilated at the lower edge}.

We now introduce some notation to parameterize these operators. Recall that $l,l'$ are topologically equivalent
if $l-l' = K \cdot \Lambda$ for some integer vector $\Lambda$. Let $\mathcal{L}$ be a set of vectors $l$ containing
one representative from each of the above equivalence classes. Let $\mathcal{M}$ be a subset of $\mathcal{L}$, consisting of
all quasiparticles that can be annihilated at the lower edge. In this notation, the most general scaling operators 
in our theory are of the form
\begin{equation}
e^{i(m+ K\Lambda)^T \phi} \mathcal{O}_{\{n_{J,k}\}} 
\end{equation}
where $m \in \mathcal{M}$, and $\Lambda$ is an integer vector.

Given this parameterization of scaling operators, the partition function $Z(\tau)$ (\ref{partfunct}) can be written as
\begin{equation}
Z(\tau) = \sum_{m \in \mathcal{M}} Z_m(\tau)
\label{partfunct2}
\end{equation}
where $Z_m$ denotes the sum (\ref{partfunct}) taken over all $\Lambda$ and $\{n_{J,k}\}$, with $m$ fixed.

To proceed further, we use the transformation law
\begin{equation}
Z_l(-1/\tau) = \sum_{l \in \mathcal{L}} S_{ll'} Z_{l'}(\tau)
\label{zltrans}
\end{equation}
where $S$ is defined by
\begin{equation}
S_{ll'} = \frac{1}{D} e^{i\theta_{ll'}} \ , \ D = \sqrt{|\det(K)|} 
\label{tops}
\end{equation}
This relation can be derived in two ways. First, it can be derived using the Poisson summation formula, as shown in 
appendix \ref{transapp}. Second, it can be established using the bulk-edge correspondence for Chern-Simons theory:
quite generally we expect the ``modular $S$-matrix'',\cite{Francesco} which is defined in terms of the edge 
partition function transformation law (\ref{zltrans}), to match the ``topological $S$-matrix'',\cite{Kitaevhoneycomb} 
which is defined in terms of the bulk quasiparticle braiding statistics (\ref{tops}).

Substituting (\ref{zltrans}) into (\ref{partfunct2}) gives
\begin{equation}
Z(-1/\tau) = \sum_{m \in \mathcal{M}, l \in \mathcal{L}} S_{ml} Z_l(\tau)
\end{equation}
Applying the modular invariance constraint (\ref{modinv}), we deduce
\begin{equation}
\sum_{m \in \mathcal{M}} Z_m(\tau) = \sum_{m \in \mathcal{M}, l \in \mathcal{L}} S_{ml} Z_l(\tau)
\label{Scond0}
\end{equation}
implying that
\begin{equation}
\sum_{m \in \mathcal{M}} S_{m l} = \begin{cases} 1 & \mbox{if }l \in \mathcal{M} \\
0 & \mbox{otherwise}
\end{cases}
\label{Scond}
\end{equation}
It is worth mentioning that there is a subtlety in deriving (\ref{Scond}) from (\ref{Scond0}). The subtlety is that 
the $\{Z_l(\tau)\}$ are not linearly independent as functions of $\tau$: in fact
$Z_l(\tau) = Z_{\wb{l}}(\tau)$ where $\wb{l} = -l$ denotes the antiparticle of $l$. However, it can be shown that
the \emph{sums} $\{Z_l(\tau) + Z_{\wb{l}}(\tau)\}$ are linearly independent as functions of $\tau$, at least for 
generic $V_{IJ}$. 
This linear independence, together with the fact that $m \in \mathcal{M}$ if and only if $\wb{m} \in \mathcal{M}$, 
allows us to deduce (\ref{Scond}) from (\ref{Scond0}).

Equation (\ref{Scond}) is the main result of this section. We now use this result to show that $\mathcal{M}$ obeys 
the two conditions from the criterion. To this end, we first consider the case where $l$ is the trivial 
quasiparticle. In this case, the right hand side of (\ref{Scond}) is $1$ while the left hand side is
\begin{equation}
\sum_{m \in \mathcal{M}} S_{ml} = \sum_{m \in \mathcal{M}} \frac{1}{D} = \frac{|\mathcal{M}|}{D}
\end{equation}
where $|\mathcal{M}|$ is the number of elements of $\mathcal{M}$. We deduce that $|\mathcal{M}| = D$. 

Next, let $l \in \mathcal{M}$ be arbitrary. In this case, the left hand side of (\ref{Scond}) is 
\begin{equation}
\sum_{m \in \mathcal{M}} S_{ml} = \sum_{m \in \mathcal{M}} \frac{e^{i\theta_{ml}}}{|\mathcal{M}|}
\end{equation}
Thus, the only way that (\ref{Scond}) can be satisfied is if all the phase factors $e^{i\theta_{ml}}$ are
equal to $1$. In other words, we must have $e^{i\theta_{m m'}} = 1$ for all $m, m' \in \mathcal{M}$. That is, 
$\mathcal{M}$ satisfies condition (1). 

Finally, we consider the case where $l \not\in \mathcal{M}$. In this case, we must have $e^{i\theta_{ml}} \neq 1$ 
for at least one $m \in \mathcal{M}$ since otherwise the left hand side of (\ref{Scond}) would evaluate to $1$ 
rather than $0$. Hence $\mathcal{M}$ must satisfy condition (2) as well. We conclude that $\mathcal{M}$ is a Lagrangian
subgroup, as claimed. 

\section{Conclusion}
In this paper, we have derived a general criterion for when an electron system with abelian quasiparticle
statistics and $K_H = 0$ can support a gapped edge: a gapped edge is possible if and only if there exists
a subset of quasiparticles $\mathcal{M} \subseteq \mathcal{L}$ with the two properties discussed in the introduction 
(i.e. a ``Lagrangian subgroup''). We have established this criterion with three arguments -- one based 
on a microscopic analysis of the edge, another on constraints from 
braiding statistics, and the third on modular invariance. 

Our analysis has also shown that every gapped edge can be associated with a Lagrangian subgroup $\mathcal{M} \subseteq \mathcal{L}$.
Physically, $\mathcal{M}$ corresponds to the set of quasiparticles that can be annihilated at the edge.
Furthermore, we have shown that there exists at least one gapped edge for each 
$\mathcal{M} \subseteq \mathcal{L}$ (appendix \ref{strongapp}). In this
sense, the different Lagrangian subgroups $\mathcal{M} \subseteq \mathcal{L}$ classify 
(or at least partially classify) the different types of gapped edges that are possible for a given bulk state. 

For concreteness, we have focused on systems built out of electrons (i.e. fermions). However, 
the criterion for a gapped edge also applies to bosonic systems with only one modification:
in the bosonic case, we require that all the quasiparticles in $\mathcal{M}$ are
\emph{bosons} in addition to the two properties from the introduction. As in the fermionic case, one can show that a
bosonic system can have a gapped edge if and only if there exists a 
``Lagrangian subgroup'' $\mathcal{M}$ with these three properties. Furthermore, the different
types of gapped edges are (at least partially) classified by the different Lagrangian subgroups $\mathcal{M} \subseteq \mathcal{L}$.
These results can be derived using arguments similar to the fermionic case, as discussed in appendix \ref{boseapp}.

Throughout this paper we have analyzed boundaries between a gapped quantum many-body system and the vacuum. More generally,
we could also consider interfaces between two gapped quantum many-body systems. Fortunately, these more
complicated geometries can be reduced to the case studied here using a simple (and well-known) trick. Specifically,
in order to understand the interface between two Hamiltonians $H, H'$, we imagine folding the system along the interface as we would fold
a sheet a paper. In this way, we can see that the $H/H'$ boundary is equivalent to an interface between the vacuum and a 
bilayer system with Hamiltonian $H + H'_r$, where $H'_r$ is obtained by a spatial reflection of $H'$. We conclude that the boundary 
between $H$ and $H'$ can be gapped if and only if the corresponding bilayer system has a Lagrangian subgroup. 

There are a number of possible directions for future work. One direction is 
to perform a more concrete analysis of protected edge modes for a particular system. For example, it 
would be interesting to investigate a specific model of the $\nu = 2/3$ edge (\ref{K23}) in the presence of 
arbitrary scattering terms (\ref{tunnel}) and explicitly verify that the edge has gapless excitations when proximity-coupled
to a superconductor. Such a calculation could also shed light on important physical properties of these edge modes such 
as their robustness to disorder and their ability to transport heat.

Another direction is to generalize the criterion to systems with \emph{non-abelian} statistics. To formulate such
a generalization, it may be helpful to study the classification of exactly soluble gapped edges given in 
Ref. \onlinecite{KitaevKong}. Other guidance may be obtained by extending the braiding statistics and modular 
invariance arguments of sections \ref{braidargsect} and \ref{modinvsect} to non-abelian states; it is less clear how to 
generalize the microscopic analysis of section \ref{microsect}.

%Finally, it would be interesting to investigate higher dimensional systems. While some three dimensional (3D) systems
%such as topological insulators have robust surface states, there are
%no known examples of 3D systems whose surface states are protected independent of any symmetry. An intriguing 
%problem is to find examples of these robust boundary modes in 3D or higher dimensional systems.
 
\acknowledgments
I would like to thank Zhenghan Wang and Maissam Barkeshli for stimulating discussions and would like to 
acknowledge support from the Alfred P. Sloan foundation.

\appendix

\section{Relation between null vectors and Lagrangian subgroups} \label{equivapp}
In this section, we show that one can find $N$ linearly independent integer vectors $\{\Lambda_1,...,\Lambda_N\}$ 
satisfying $\Lambda_i^T K \Lambda_j = 0$ if and only if there exists a set of (inequivalent) integer vectors $\mathcal{M}$ satisfying 
two properties: 
\begin{enumerate}
\item{$m^T K^{-1} m'$ is an integer for any $m, m' \in \mathcal{M}$.}
\item{If $l$ is not equivalent to any element of $\mathcal{M}$, then $m^T K^{-1} l$ is non-integer for some $m \in \mathcal{M}$.}
\end{enumerate}
Here $K$ is a $2N \times 2N$ symmetric integer matrix with vanishing signature, non-vanishing determinant, and at 
least one odd element on the diagonal. 

We prove this result in section \ref{proofapp}; we then explain its physical interpretation in section \ref{physintapp}, and
we state and prove a sharper version of this result in section \ref{strongapp}. We derive a bosonic analogue in appendix 
\ref{boseapp}.

\subsection{Proof} \label{proofapp}
We first establish the ``only if'' direction. Suppose that one can find $N$ linearly independent integer vectors
$\{\Lambda_1,...,\Lambda_n\}$ such that $\Lambda_i^T K \Lambda_j = 0$. We wish to construct a set $\mathcal{M}$ of integer
vectors satisfying the two properties listed above. The first step is to make an (integer) change of basis so that 
the last $N$ components of $\Lambda_i$ are all zero for every $\Lambda_i$. In the new basis, the matrix $K$ has the block diagonal form
\begin{equation}
K = \bpm 0 & A \\ A^T & B \epm
\end{equation}
where $A,B$ are $N \times N$ matrices. Hence, $K^{-1}$ is given by
\begin{equation}
K^{-1} = \bpm -(A^T)^{-1} B A^{-1} & (A^T)^{-1} \\ A^{-1} & 0 \epm
\end{equation}

We then let $\mathcal{M}$ be the set of all vectors of the form $\bpm 0 \\ v \epm$ where $v$ is an $N$ component integer vector. 
(More precisely, we divide this set into equivalence classes modulo $K \mathbb{Z}^{2N}$, and choose one vector 
from each equivalence class).

We can easily see that $\mathcal{M}$ satisfies the two properties listed above. To establish the first property,
note that $m^T K^{-1} m' = 0$ for any $m, m' \in \mathcal{M}$, so in particular this quantity is always an integer.
As for the second property, let $l = \bpm u_1 \\ u_2 \epm $ be an integer 
vector such that $l^T K^{-1} m$ is an integer for all $m \in \mathcal{M}$. Then $u_1 = A w $ for some integer vector $w$,
so we can write 
\begin{equation}
l = K \cdot \bpm 0 \\ w \epm + \bpm 0 \\ u_2 - B w \epm
\end{equation}
Examining this expression, we see that $l$ is equivalent to an element of $\mathcal{M}$. This is what we wanted to show.

We next establish the ``if'' direction. Suppose $\mathcal{M}$ is a set of vectors satisfying the above two properties.
We wish to construct $\{\Lambda_1,...,\Lambda_N\}$ satisfying $\Lambda_i^T K \Lambda_j = 0$. To this end, let us 
consider the set 
\begin{equation}
\Gamma = \{m + K \Lambda\ : m \in \mathcal{M}, \Lambda \in \mathbb{Z}^{2N}\}
\label{gammadef}
\end{equation}
This set forms a $2N$ dimensional 
integer lattice, and therefore can be represented as $\Gamma = U \mathbb{Z}^{2N}$ where $U$ is some $2N \times 2N$ integer matrix.

Now consider the matrix $P = U^T K^{-1} U$. We claim that $P$ is a symmetric integer matrix with vanishing signature, 
determinant $\pm 1$, and at least one odd element on the diagonal. Indeed, the fact 
that $P$ is symmetric, has vanishing signature, and has at least one odd element on the diagonal, follows from the corresponding 
properties of $K$. Also, the fact that $P$ is an integer matrix follows from the first property of $\mathcal{M}$. 
Finally, to see that $P$ has determinant $\pm 1$, we use the second property of $\mathcal{M}$: 
we note that if $x \not\in \mathbb{Z}^{2N}$, then $y^T P x$ is non-integer for some $y \in \mathbb{Z}^{2N}$. 
Hence, if $x \not\in \mathbb{Z}^{2N}$, then $P x \not\in \mathbb{Z}^{2N}$. It follows that $P^{-1}$ must be an integer 
matrix, so that $P$ has determinant $\pm 1$. 

The next step is to use the following theorem, due to Milnor: \cite{MilnorBook} suppose $A,A'$ are two symmetric, indefinite, 
integer matrices with determinant $\pm 1$. Suppose in addition that $A,A'$ have the same dimension and same signature and
are either both even or both odd -- where an ``even'' matrix has only even elements on the diagonal, and an ``odd'' matrix 
has at least one odd 
element on the diagonal. Milnor's theorem (Ref. \onlinecite{MilnorBook}, p. 25) states that there must exist an integer 
matrix $W$ with unit determinant such that $W^T A W = A'$. 

Applying this result to the matrix $P$ (an ``odd'' matrix with vanishing signature) we deduce that we can always block 
diagonalize $P$ as
\begin{equation}
W^T P W = \bpm \bf{1} & 0 \\ 0 & -\bf{1} \epm
\label{Peq}
\end{equation}
where $W$ is an integer matrix with $\det(W) = \pm 1$ and $\bf{1}$ denotes the $N \times N$ identity matrix. 

To complete the argument, we define $v_i = w_i + w_{i+N}$ where $w_i$ is the $i$th column of $W$.
We then define 
\begin{equation}
\Lambda_i = \det(K) \cdot K^{-1} U v_i
\label{lambdadef}
\end{equation}
It is easy to check that the $\Lambda_i$ obey $\Lambda_i^T K \Lambda_j = 0$, and are linearly independent 
and integer. 

\subsection{Understanding the correspondence} \label{physintapp}
The ``only if'' part the argument shows that every collection of null vectors $\{\Lambda_i\}$ can be
associated with a corresponding Lagrangian subgroup $\mathcal{M}$. We now discuss the physical meaning
of this $\{\Lambda_i\} \rightarrow \mathcal{M}$ correspondence and show that it agrees with the physical 
picture of section \ref{braidargsect}. 

To begin, it is helpful to reformulate the correspondence in a basis independent 
way: given any linearly independent $\{\Lambda_1,...,\Lambda_N\}$ satisfying $\Lambda_i^T K \Lambda_j = 0$, 
we define $\mathcal{M}$ to be the set of all (inequivalent) vectors $m \in \mathbb{Z}^{2N}$ such that
\begin{equation}
b \cdot m =  \sum_i a_i \cdot K \Lambda_i
\label{Mdef}
\end{equation}
for some $b, a_i \in \mathbb{Z}$. It is easy to verify that this definition of $\mathcal{M}$ agrees with 
the one given in the previous section.

This alternative formulation is useful because it reveals the physical interpretation of the 
$\{\Lambda_i\} \rightarrow \mathcal{M}$ correspondence: the set $\mathcal{M}$ is
simply the set of quasiparticles that can be annihilated at the gapped edge corresponding to $\{\Lambda_i\}$.
To see this, note that when $\sum_i U(\Lambda_i)$ gaps the edge, it freezes the value of $\Lambda_i^T K \phi$ 
and hence also freezes the value of the linear combination $\sum_i a_i \Lambda_i^T K \phi$. It then follows from (\ref{Mdef}) 
that the value of $m^T \phi$ is frozen for each $m \in \mathcal{M}$. Thus, we expect the operator $e^{i m^T \phi}$ to exhibit long range
order: 
\begin{equation}
\<e^{i m^T \phi(x_1)} e^{-im^T \phi(x_2)}\> = \text{const.} \neq 0
\end{equation}
in the limit $|x_1 - x_2| \rightarrow \infty$. 
This long rang order implies that the associated quasiparticle $m$ can be annihilated at the edge. Indeed, according to the bulk-edge 
correspondence,\cite{WenReview,WenBook} the operator $e^{i m^T (\phi(x_1) - \phi(x_2))}$ can be interpreted as 
describing a process in which two quasiparticles $m, \wb{m}$ are created in the bulk and brought to points $x_1, x_2$ at the edge. 
Hence, $\<e^{i m^T \phi(x_1)} e^{-im^T \phi(x_2)}\>$ can be thought of as an overlap between the group state $|\Psi\>$, and an excited state 
$|\Psi_{ex}\> = e^{i m^T \phi(x_1)} e^{-im^T \phi(x_2)}|\Psi\>$ with two quasiparticles at the edge. The fact that this overlap is nonzero 
implies that the corresponding quasiparticles $m,\wb{m}$ can be annihilated at the edge, as shown in Lemma 1 of appendix \ref{prelimapp}. 

\subsection{Sharpening the correspondence} \label{strongapp}
While the above argument shows that every state with at least one Lagrangian subgroup $\mathcal{M} \subseteq \mathcal{L}$ can 
support at least one type of gapped edge, some states can have more than one Lagrangian subgroup. Thus, it is desirable to prove a 
stronger result -- namely, every Lagrangian subgroup $\mathcal{M}$ can be associated with a \emph{corresponding} gapped edge. 
Such a result, together with our proof that every gapped edge is associated with a Lagrangian subgroup, would imply that the 
different types of gapped edges are (at least partially) classified by Lagrangian subgroups $\mathcal{M} \subseteq \mathcal{L}$. 

We now derive this sharper result. That is, we construct a gapped edge for each $\mathcal{M}$ in such a way that 
\emph{the quasiparticles in $\mathcal{M}$ can be annihilated at the boundary}. We would like to mention that while this paper was 
being revised to include this extension of appendix \ref{proofapp}, we became aware that Barkeshli, Jian, and Qi, 
making use of an earlier draft of this paper, have obtained a similar extension.\cite{Barkeshliunpub}

To prove this stronger result, we modify our previous construction of $\{\Lambda_1,...,\Lambda_N\}$ 
(\ref{lambdadef}). The most important modification is that we use a more complicated edge theory: instead of considering 
the standard edge theory (\ref{edgeth}) for the Chern-Simons theory (\ref{cstheory}), we consider an enlarged edge theory 
with a $4N \times 4N$ $K$-matrix 
\begin{equation}
K' = \bpm K & 0 & 0 \\
	  0 & -\mathbf{1} & 0 \\
	  0 &  0 & \mathbf{1} \epm
\label{kprime}
\end{equation}
Here, $\mathbf{1}$ is an $N \times N$ identity matrix. Physically, the $K'$ edge theory can be realized in an edge reconstruction scenario where 
$N$ non-chiral Luttinger liquids, described by $\bpm 1 & 0 \\ 0 & -1 \epm$ are glued to the standard edge for $K$. 

We next describe how to construct null vectors $\{\Lambda_i\}$ that gap out the $K'$ edge \emph{and} give us a boundary where the particles
in $\mathcal{M}$ can be annihilated. Since the $K'$ edge has $4N$ chiral modes, we need
$2N$ vectors $\{\Lambda_1,...,\Lambda_{2N}\}$ with $\Lambda_i^T K' \Lambda_j = 0$. We construct $\{\Lambda_i\}$ using
the same recipe as above. First, we define a $2N$ dimensional lattice $\Gamma$ by equation (\ref{gammadef}) 
and we construct a matrix $U$ with $\Gamma = U \mathbb{Z}^{2N}$. We then define $P = U^T K^{-1} U$, and we find a unit determinant matrix $W$ satisfying
equation (\ref{Peq}). The only new element comes in the definition of $\Lambda_i$. In the modified construction, we set
\begin{equation}
\Lambda_i = \det{K} \cdot \bpm K^{-1} U w_i \\ e_i \epm
\end{equation}
where $w_i$ is the $i$th column of $W$ and $e_i$ is a $2N$ component vector with a $1$ in the $i$th entry, and all other entries vanishing.
It is easy to verify that $\Lambda_i^T K' \Lambda_j = 0$ and that the $\{\Lambda_i\}$ are all integer vectors.

At this point, it is clear that the perturbation $\sum_{i=1}^{2N} U(\Lambda_i)$ will gap the $K'$ edge. All that remains is to prove that
the quasiparticles in $\mathcal{M}$ can be annihilated at this edge. To this end, we note that
\begin{align} 
K' \Lambda_i = \det{K} \cdot \bpm U w_i \\ -\Sigma_z e_i \epm \ \ , \ \ \Sigma_z = \bpm \mathbf{1} & 0 \\ 0 & -\mathbf{1} \epm
\label{kprimeeq}
\end{align}
We then recall that $W$ has unit determinant, so the lattice generated by $\{U w_i\}$ spans all of $U \mathbb{Z}^{2N} = \Gamma$. Since $\Gamma$
contains every $m \in \mathcal{M}$, we can see from (\ref{kprimeeq}) that the lattice generated by $\{K' \Lambda_i\}$ contains  
the vector $\det(K) \cdot m$ for every $m \in \mathcal{M}$ (modulo $K' \mathbb{Z}^{4N}$). Applying the analysis of section \ref{physintapp}
we conclude that all the quasiparticles in 
$\mathcal{M}$ can be annihilated at the boundary. 

\section{Proof that the null vector criterion is necessary} \label{nullnecapp}
In this section, we consider the two component edge theory (\ref{edgeth}) in the presence of 
a single scattering term $U(\Lambda)$ (\ref{tunnel}). We show that a necessary condition for $U(\Lambda)$ 
to gap the edge is that $\Lambda$ satisfy the null vector criterion, $\Lambda^T K \Lambda = 0$.

Our basic strategy is to construct a (fictitious) $U(1)$ charge $Q$ which is conserved by $U(\Lambda)$, and then show 
that the system has a nonzero Hall conductivity with respect to this charge. To this end, we consider a general
$U(1)$ charge of the form
\begin{equation}
Q = \frac{1}{2\pi} \int t^T \partial_x \phi
\label{chargedef}
\end{equation}
where $t^T = (t_1, t_2)$ is some two component real vector. 

Next, we choose $t_1 = b, \ t_2 = -a$  where $\Lambda = (a,b)$. This choice of $t$ guarantees that 
\begin{equation}
[Q, U(\Lambda)] = 0
\end{equation}
As a result, the charge $Q$ is a conserved 
quantity so it is sensible to compute the associated Hall conductivity $\sigma_{H}^Q$. 
Following the usual $K$-matrix formalism we have:
\begin{eqnarray}
\sigma_{H}^{Q} &=& t^T K^{-1} t \nonumber \\
&=& \frac{1}{\det(K)} \bpm b & -a \epm \cdot \bpm K_{22} & -K_{12} \\ -K_{21} & K_{11} \epm \cdot
\bpm b \\ -a \epm \nonumber \\
&=& \frac{1}{\det(K)} (K_{11} \cdot a^2 + 2 K_{12} \cdot a b + K_{22} \cdot b^2) \nonumber \\
&=& \frac{1}{\det(K)} \Lambda^T K \Lambda
\end{eqnarray}

We are now finished: we can see that if $\Lambda$ doesn't satisfy the null vector criterion (\ref{null}), then
$\sigma_{H}^{Q} \neq 0$. It then follows that $U(\Lambda)$ cannot gap out the edge, since a system with a
nonzero Hall conductivity has a protected edge if the corresponding charge is conserved. \cite{Laughlinflux,Halperinedge}
This proves the claim.

We would like to emphasize that the above argument does \emph{not} rule out the possibility of gapping the edge
with other types of perturbations. In fact, it does not even rule out simple perturbations
like a sum of \emph{two} scattering terms $U(\Lambda_1)+U(\Lambda_2)$: these terms break all
the $U(1)$ symmetries at the edge, thus invalidating the above analysis. 

\section{Establishing condition (2) of the criterion} \label{braidnondegapp}
In this section, we consider a general gapped electron system which has abelian quasiparticle statistics and has a 
gapped edge. For this class of systems, we argue that the set of quasiparticles 
that can be annihilated at the edge (denoted by $\mathcal{M}$) must obey condition (2) of the criterion. In other words, 
we show that if $l$ is a quasiparticle that cannot be annihilated at a gapped edge, then $l$ must have
nontrivial statistics with at least one quasiparticle $m$ that can be annihilated at the edge. 

The argument we present is not a rigorous mathematical proof: we do not give precise definitions for
all the concepts that we use, and we regularly drop quantities that we expect to vanish in the thermodynamic
limit. Despite these limitations, we believe that the argument could be used as a starting point 
for constructing a rigorous proof.

\subsection{Preliminaries} \label{prelimapp}
Our argument relies on the following conjecture about gapped many-body systems:

\vspace{2 mm}

{\bf Conjecture 1:} {\it Let $|\Psi\>$ be the ground state of a 2D gapped many-body system defined in a spherical geometry. 
Let $|\Psi'\>$ be another state (not necessarily an eigenstate) which has the same energy density outside of 
two non-overlapping disk-like regions $A,B$. Then we can write 
\begin{equation}
|\Psi'\> = \sum_k U_k W_k |\Psi\>
\label{conjeq}
\end{equation}
where $W_k$ is a (string-like) unitary operator that describes a process in which a pair of quasiparticles 
$k, \wb{k}$ are created and then moved to regions $A,B$ respectively, and where $U_k$ is an operator acting 
within $A \cup B$. Here, the sum runs over different quasiparticle types $k$.}

\vspace{2 mm}

In more physical language, the above conjecture is the statement that any excited state whose
excitations are located in two disconnected regions $A,B$ can be constructed 
by moving a pair of quasiparticles $k, \wb{k}$ into $A,B$ and then applying an operator $U_k$
acting within $A \cup B$. This claim is reasonable because we expect that the different excited
states of a gapped many-body system can be divided into topological sectors parameterized by the quasiparticle 
type $k$, and that any two excitations in the same sector can be transformed into one another by local operations.

In addition, we will make use of the following lemma:

\vspace{2mm}

{\bf Lemma 1:} {\it Consider a 2D gapped many-body system with a gapped edge. Let $|\Psi\>$ denote the ground state and let 
$|\Psi_{ex}\>$ denote an excited state with a quasiparticle $l$ and quasihole $\wb{l}$ located near two
points $a,b$ at the boundary. If $l$ cannot be annihilated at the edge then
\begin{equation}
\lim_{|a-b| \rightarrow \infty} \<\Psi | U_{a} U_{b} | \Psi_{ex}\> = 0
\label{lemmaeq}
\end{equation}
for any operators $U_a$, $U_b$ acting near $a,b$.}

\vspace{2 mm}

To derive this result, let $H$ be a gapped, local Hamiltonian whose ground state is $|\Psi\>$.
Let $H_{ex}$ be a gapped, local Hamiltonian whose ground state is $|\Psi_{ex}\>$. We can assume without loss of
generality that the ground state energies of $H, H_{ex}$ are both $0$: 
\begin{equation}
H |\Psi\> = H_{ex}|\Psi_{ex}\> = 0
\end{equation} 
We will also assume that $H_{ex}$ can be written as 
\begin{equation}
H_{ex} = H + H_a + H_b
\end{equation}
 where $H_a, H_b$ are local operators acting near $a,b$. 

We will now show that if
\begin{equation}
\lim_{|a-b| \rightarrow \infty} \< \Psi | U_a U_b |\Psi_{ex}\> = \alpha \neq 0
\end{equation}
then we can always construct ``dressed'' operators $\mathcal{U}_a, \mathcal{U}_b$ such that
$\lim_{|a-b| \rightarrow \infty} \mathcal{U}_a \mathcal{U}_b |\Psi_{ex}\> = |\Psi\>$. This will establish the 
lemma (since the latter equation means that $l$ can be annihilated at the edge).

To do this, we use a trick due to Hastings (Ref. \onlinecite{Hastingsloc}) and Kitaev (Ref. \onlinecite{Kitaevhoneycomb}, 
appendix D.1.2). Let $\tilde{f}(\omega)$ be a real, smooth function satisfying
\begin{align}
\tilde{f}(0) = 1 \ , \ \tilde{f}(\omega) =0 \ \text{for} \ |\omega| \geq \Delta
\label{fprop}
\end{align}
where $\Delta$ is the energy gap of $H$. Define 
\begin{equation}
f(t) = \frac{1}{2\pi}\int_{-\infty}^{\infty} d\omega \tilde{f}(\omega) e^{-i \omega t}
\label{fdef}
\end{equation}
Given that $\tilde{f}(\omega)$ is smooth, it follows that $f(t) \rightarrow 0$ as $t \rightarrow \infty$ faster than any polynomial. We then define
\begin{equation}
\mathcal{U} = \frac{1}{\alpha}\int_{-\infty}^{\infty} dt \ f(t) \cdot e^{i Ht} U_a U_b e^{-i H_{ex}t}
\end{equation} 

Straightforward algebra gives
\begin{eqnarray}
\mathcal{U} |\Psi_{ex}\> &=& \frac{1}{\alpha}\int_{-\infty}^{\infty} dt \ f(t) \cdot e^{i Ht} U_a U_b |\Psi_{ex}\> \nonumber \\
&=& \frac{1}{\alpha}\int_{-\infty}^{\infty} dt \ f(t) \cdot \sum_n e^{i E_n t} |\Psi_n\>\<\Psi_n | U_a U_b |\Psi_{ex}\> \nonumber \\
&=& \frac{1}{\alpha} |\Psi\>\<\Psi | U_a U_b |\Psi_{ex}\> 
\end{eqnarray}
so that
\begin{equation}
\lim_{|a-b| \rightarrow \infty} \mathcal{U} |\Psi_{ex}\> = |\Psi\>
\end{equation}
Furthermore, since $f$ decays rapidly as $t \rightarrow \infty$, and $H$ is a local Hamiltonian, it is not hard to see that 
the region of support of $\mathcal{U}$ is well-localized near $a$ and $b$. Also, $\mathcal{U}$ can be (approximately) factored as 
$\mathcal{U} = \mathcal{U}_a \cdot \mathcal{U}_b$, up to terms that decay rapidly in the separation between $a$ and $b$. In this
way, we can explicitly construct operators $\mathcal{U}_a, \mathcal{U}_b$ acting near $a,b$ such that 
$\lim_{|a-b| \rightarrow \infty} \mathcal{U}_a \mathcal{U}_b |\Psi_{ex}\> = |\Psi\>$.

\begin{figure}[tb]
\centerline{
\includegraphics[width=0.99\columnwidth]{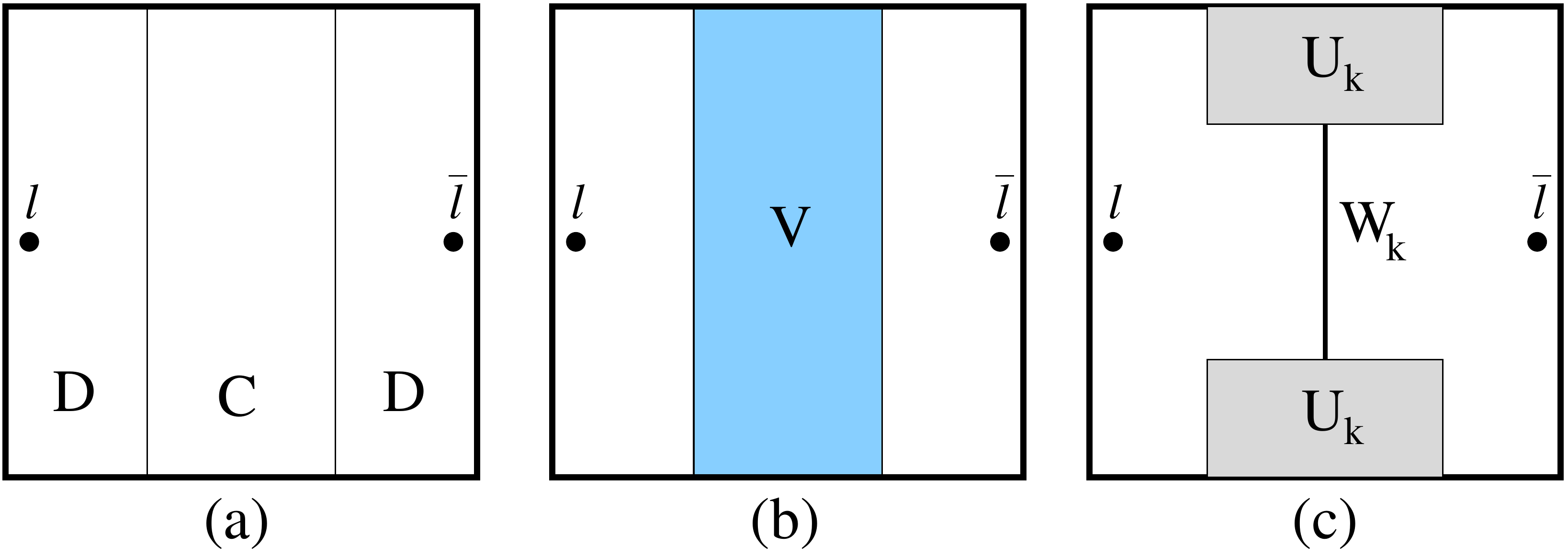}
}
\caption{
Key steps in the argument.
(a) We partition the $L \times L$ system into two pieces $C,D$ where $D$ has two connected components --
one containing $l$ and one containing $\wb{l}$.
(b) We show that there exists an operator $V$ acting in region $C$ such that $V|\Psi\> = |\Psi\>$
and $V|\Psi_{ex}\> = 0$.
(c) We show that we can replace $V$ by an operator of the form $\sum_k U_k W_k$
where $U_k$ are operators acting near the boundary, and $W_k$ are (string-like) unitary operators that describe a 
process in which a pair of quasiparticles $k, \bar{k}$ are created in the bulk and
moved near the boundary. 
}
\label{vfig}
\end{figure}

\subsection{Outline of argument}
We consider a system in a square $L \times L$ geometry with a gapped edge. We let $|\Psi\>$ denote the 
ground state, and let $|\Psi_{ex}\>$ denote a state with two quasiparticles $l, \wb{l}$ located near two
well-separated points at the boundary (Fig. \ref{vfig}a). The argument proceeds in three steps:
\begin{enumerate}
\item{
In the limit $L \rightarrow \infty$, we show that there exists an operator $V$ acting in region $C$ such 
that $V |\Psi\> = |\Psi\>$ and $V |\Psi_{ex}\>= 0$ (Fig. \ref{vfig}b). 
}
\item{
We show that we can replace $V$ by an operator of the form $\sum_k U_k W_k$
where $U_k$ are operators acting near the boundary, and $W_k$ are (string-like) unitary operators that describe a 
process in which a pair of quasiparticles $k, \bar{k}$ are created in the bulk and
moved near the boundary (Fig. \ref{vfig}c). That is, we show that
$\sum_k U_k W_k|\Psi\> = |\Psi\>$ and $\sum_k U_k W_k |\Psi_{ex}\> = 0$.
}
\item{
We show that there is at least one quasiparticle $k$ that has nontrivial statistics with respect to $l$ and can be
annihilated at the edge. This result proves the claim: the set $\mathcal{M}$ of quasiparticles that can be annihilated
at the edge obeys condition (2) of the criterion.
}
\end{enumerate}

\subsection{Step 1}
The first step is to partition our system into two pieces, $C$ and $D$ (Fig. \ref{vfig}a). We note that $D$ has two
connected components -- one containing $l$ and one containing $\wb{l}$. We will now show that in the limit 
$L \rightarrow \infty$ there exists an operator $V$ acting in region $C$ with $V |\Psi\> = |\Psi\>$ and 
$V |\Psi_{ex}\> = 0$ (Fig. \ref{vfig}b). 

To construct $V$, we consider the Schmidt decomposition of $|\Psi\>$ corresponding to the bipartition $C,D$:
\begin{equation}
|\Psi\> = \sum_i \lambda_i |\Psi_{C,i}\> \otimes |\Psi_{D,i}\> 
\end{equation}
Here $\{|\Psi_{C,i}\>\}$ and $\{|\Psi_{D,i}\>\}$ are orthonormal many-body states corresponding to regions $C$ and $D$ 
and $\lambda_i$ are Schmidt coefficients. Since the $\{|\Psi_{C,i}\>\}$, $\{|\Psi_{D,i}\>\}$ form a complete 
orthonormal basis for $C$ and $D$, we can also express $|\Psi_{ex}\>$ in terms of these states: 
\begin{equation}
|\Psi_{ex}\> = \sum_{ij} \lambda_{ij}' |\Psi_{C,i}\> \otimes |\Psi_{D,j}\>
\end{equation}
We next observe that, in the limit $L \rightarrow \infty$, the coefficients have the property that for each $i$, 
either (1) $\lambda_i = 0$, or (2) all the $\{\lambda'_{ij}\}$ vanish simultaneously.
Indeed, if $\lambda_i$ and $\lambda'_{ij}$ were both nonzero for some $i,j$ then the operator
$|\Psi_{D,i}\>\<\Psi_{D,j}|$ would have a nonzero matrix element between $|\Psi\>$ and $|\Psi_{ex}\>$. But such a nonzero matrix
element is not possible according to Lemma 1 (\ref{lemmaeq}), since $|\Psi_{D,i}\>\<\Psi_{D,j}|$ is a local operator acting 
in the region $D$.

Given this observation, we can now construct the desired operator $V$. We define 
\begin{equation}
V = \sum_{\lambda_i \neq 0} |\Psi_{C,i}\> \<\Psi_{C,i}|
\end{equation}
By construction, we have 
\begin{align}
V |\Psi\> = |\Psi\> \ , \ V |\Psi_{ex}\> = 0
\label{Veq}
\end{align}
as required. 

\begin{figure}[tb]
\centerline{
\includegraphics[width=0.9\columnwidth]{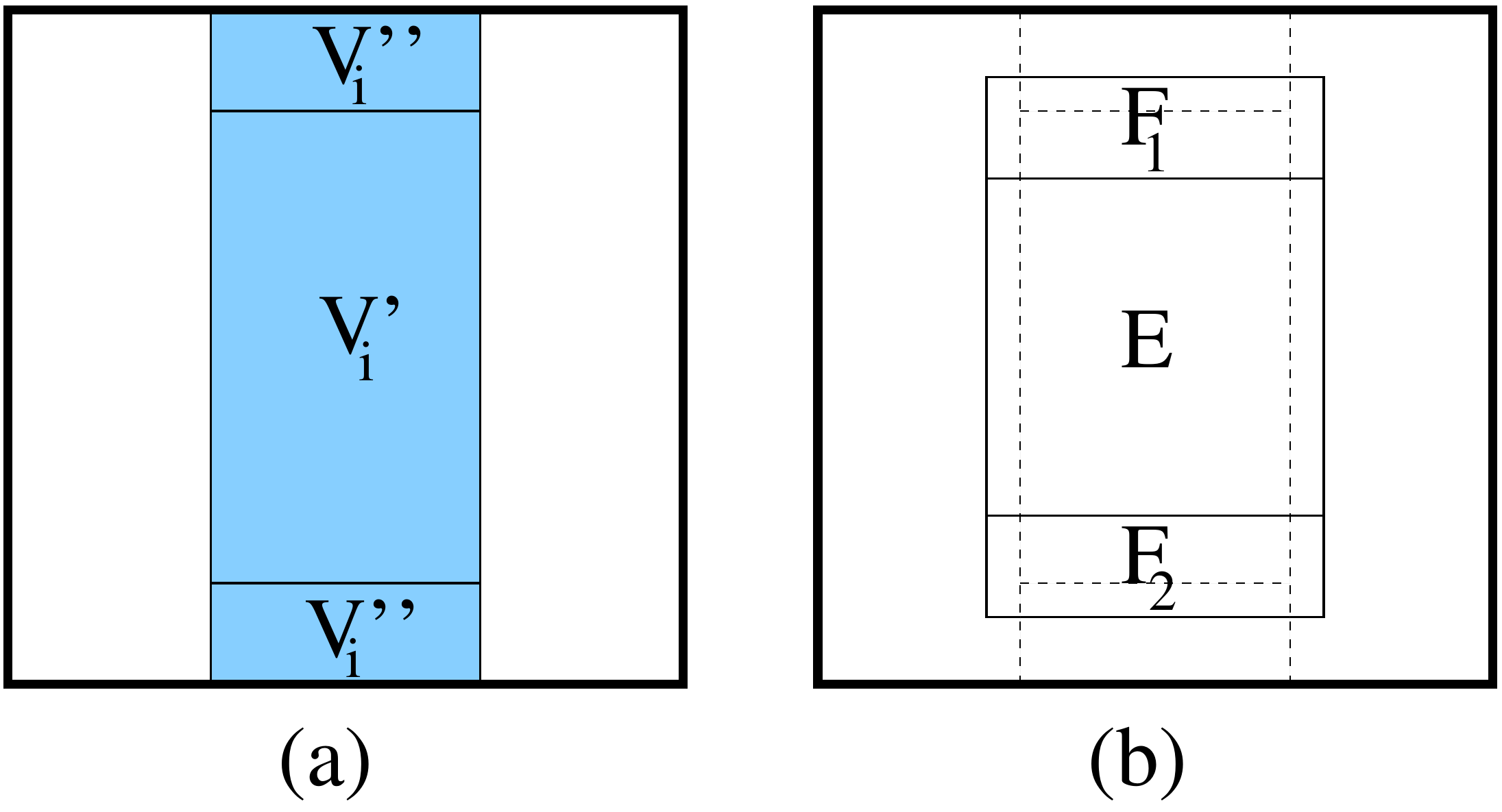}
}
\caption{
(a) We write $V =\sum_i V_i' \cdot V_i''$ where $V_i'$ acts in the interior of the system and $V_i''$ acts near the boundary.
(b) We construct an operator $P$ that acts within the region $E$, and satisfies several properties. First $P |\Psi\> =|\Psi\>$.
Second, if $P$ is applied to a state with no excitations outside of $E \cup F_1 \cup F_2$, it returns a state with no 
excitations outside of $F_1 \cup F_2$. }
\label{pfig}
\end{figure}

\subsection{Step 2} 
To proceed further, we decompose $V$ as  
\begin{equation}
V = \sum_i V''_i \cdot V'_i
\end{equation}
where $V'_i$ acts in the interior of the system and $V''_i$ acts near the two boundaries (Fig. \ref{pfig}a).

Next, we construct an operator $P$ that acts within 
the region $E$ shown in Figure \ref{pfig}b and satisfies several properties. First, $P|\Psi\> = |\Psi\>$. Second, if $P$ is applied to
a state that has no excitations outside of $E \cup F_1 \cup F_2$, it returns a state with no excitations outside of
$F_1 \cup F_2$. In other words, $P$ projects out any excitations within $E$. 

It is easy to construct $P$ in the case where the Hamiltonian $H$ is a sum of local, commuting projectors, 
$H = -\sum_i P_i$: in that case, the operator $P = \prod_{i \in E} P_i$ satisfies all the required conditions. 
In the general case, we need to work a bit harder. Let $H$ be a gapped Hamiltonian whose ground state is $|\Psi\>$. 
We can write 
\begin{equation}
H = H_E + H_F + H_0
\end{equation}
 where $H_E$ contains terms acting in (or near) region $E$, $H_F$ contains terms 
acting in (or near) region $F_1 \cup F_2$, and $H_0$ contains all the other terms in the Hamiltonian. In general, 
$H_E, H_F, H_0$ may not commute with one another since they may overlap along the boundaries between the various 
regions. However, according to a result from Ref. \onlinecite{Hastingsloc} as well as Ref. \onlinecite{Kitaevhoneycomb}, 
appendix D.1.2, we 
can always choose $H_E, H_F, H_0$ so that $|\Psi\>$ is a simultaneous eigenstate of all three operators: 
\begin{equation}
H_0 |\Psi\> = H_E |\Psi\> = H_F |\Psi\> = 0
\end{equation} 
To proceed further, we use the same trick as in the proof of Lemma 1. We choose
a real, smooth function $\tilde{f}$ satisfying (\ref{fprop}) where $\Delta$ denotes 
the bulk gap of the Hamiltonian $H_0 + H_E$. We then construct the Fourier transform $f$ (\ref{fdef})
and define
\begin{equation}
P = \int_{-\infty}^{\infty} dt \ f(t) \cdot e^{i(H_0 + H_E)t} e^{-i H_0 t}
\end{equation}
In the same way as in Lemma 1, one can verify that $P$ has all of the required properties.

\begin{figure}[tb]
\centerline{
\includegraphics[width=0.9\columnwidth]{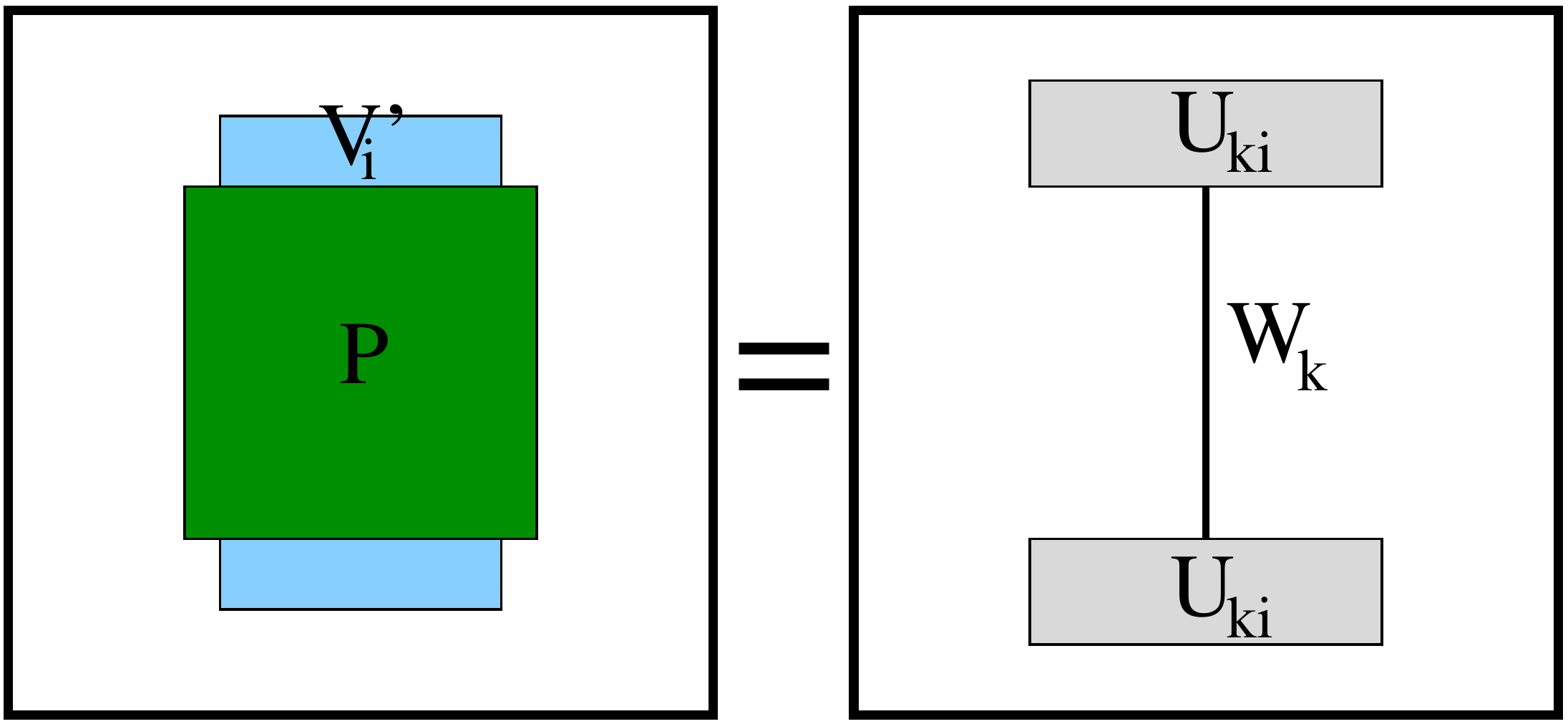}
}
\caption{
According to Conjecture 1 (\ref{conjeq}), we can write $PV'_i |\Psi\> = \sum_k U_{ki} W_k |\Psi\>$, 
where $W_k$ is a (string-like) unitary operator that describes a process in which a pair of quasiparticles 
$k, \wb{k}$ are created and then moved to regions $F_1, F_2$ (Fig. \ref{pfig}b), and
where $U_{ki}$ is an operator acting within $F_1 \cup F_2$. 
}
\label{pvfig}
\end{figure}

Now consider the state $V'_i |\Psi\>$. This state has no excitations outside of $E \cup F_1 \cup F_2$
since $V_i'$ acts entirely within this region (Fig. \ref{pfig}b). It follows that the state $P V'_i |\Psi\>$ 
has no excitations outside of $F_1 \cup F_2$. Therefore, according to Conjecture 1 (\ref{conjeq}), we can write
\begin{equation}
PV'_i |\Psi\> = \sum_k U_{ki} W_k |\Psi\>
\label{pveq1}
\end{equation}
where $W_k$ is a (string-like) unitary operator that describes a process in which a pair of quasiparticles 
$k, \wb{k}$ are created and then moved to regions $F_1, F_2$ respectively, and
where $U_{ki}$ is an operator acting within $F_1 \cup F_2$ (Fig. \ref{pvfig}). Here the sum runs over 
different particle types $k$. 

\begin{figure}[tb]
\centerline{
\includegraphics[width=0.95\columnwidth]{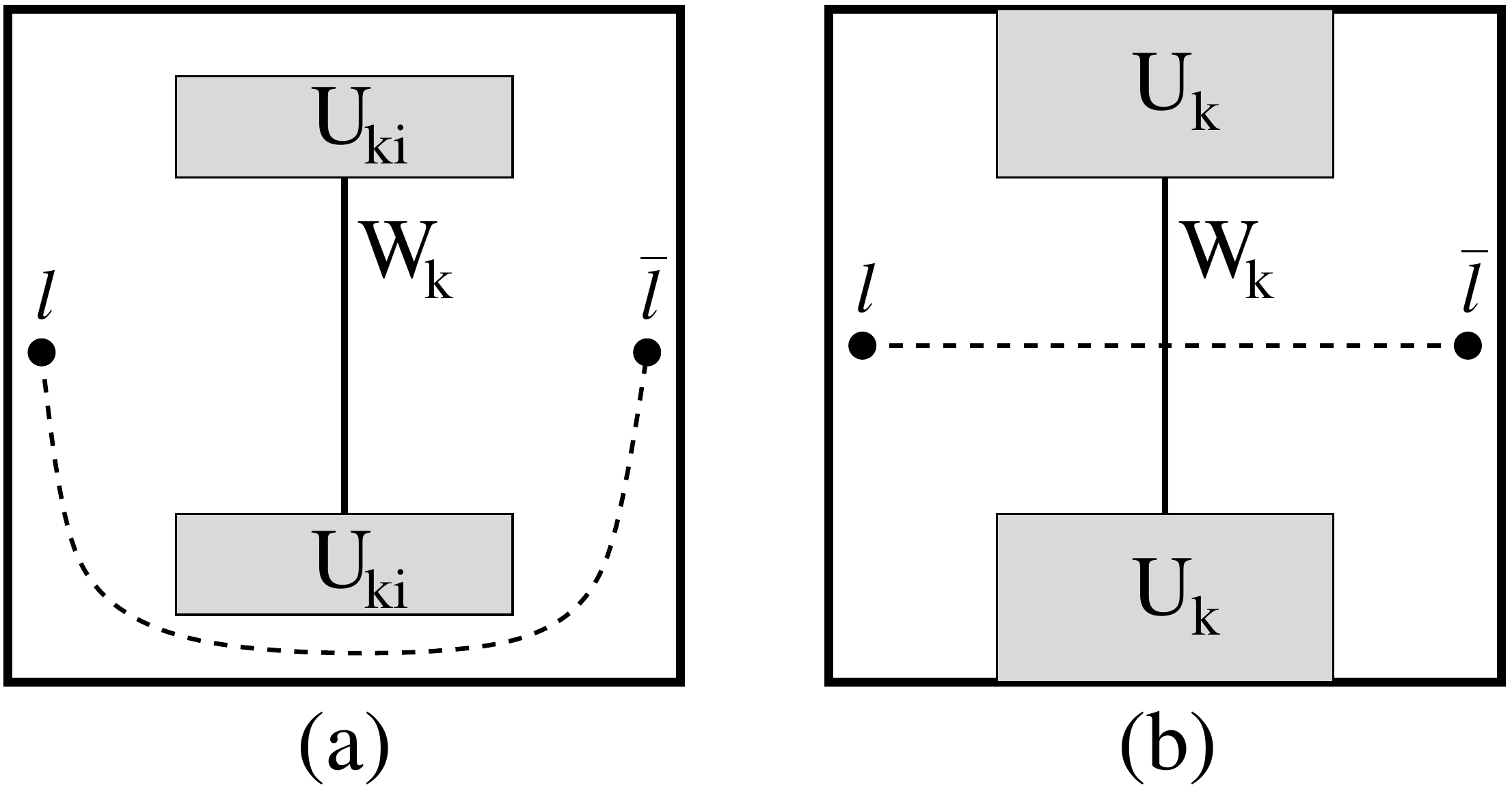}
}
\caption{
The state $|\Psi_{ex}\>$ can be written as $|\Psi_{ex}\> = W_{l\gamma} |\Psi\>$ where
$W_{l \gamma}$ is a string-like operator with path $\gamma$. This equation holds for
any choice of $\gamma$. (a) To prove (\ref{pveq2}), we choose $\gamma$ (dotted line) so that it avoids the region of support
of $\sum_k U_{ki} W_k$ and $PV_i'$. (b) To prove (\ref{braid}), we choose $\gamma$ so that
it intersects $W_k$.
}
\label{wlgammafig}
\end{figure}

We next argue that the same relation holds for $|\Psi_{ex}\>$:
\begin{equation}
P V'_i |\Psi_{ex}\> = \sum_k U_{ki} W_k |\Psi_{ex}\>
\label{pveq2}
\end{equation}
To see this, note that 
\begin{equation}
|\Psi_{ex}\> = W_{l\gamma} |\Psi\>
\label{psiexeq}
\end{equation}
where $W_{l\gamma}$ is a unitary operator that describes a 
process in which two quasiparticles $l, \wb{l}$ are created in the bulk and moved along a path $\gamma$ to 
the boundary. Furthermore, this equation holds for \emph{any} path $\gamma$ with endpoints located at
the correct positions; we are free to choose the path $\gamma$ to make our life as easy as possible. Here, we
choose $\gamma$ so that it avoids the region of support of the operators $PV_i'$ and $\sum_k U_{ki} W_k$
(Fig. \ref{wlgammafig}a).
It is then clear that $W_{l\gamma}$ commutes with $P V_i'$ and $\sum_k U_{ki} W_k$. If we then multiply both
sides of equation (\ref{pveq1}) by $W_{l\gamma}$, and commute the operators on both sides, 
the claim (\ref{pveq2}) follows immediately.

To complete Step 2 of the argument, we define 
\begin{equation}
U_k = \sum_i V''_i U_{ki}
\end{equation}
We then note that
\begin{eqnarray}
\sum_k U_k W_k |\Psi\> &=& \sum_{ik} V''_i U_{ki} W_k |\Psi\> \nonumber \\
&=& \sum_i V''_i (P V'_i |\Psi\>) \nonumber \\
&=& P \sum_i V''_i V'_i |\Psi\> \nonumber \\
&=& P V |\Psi\> \nonumber \\
&=& |\Psi\>
\label{ukwk1}
\end{eqnarray}
By the same reasoning, we have
\begin{equation}
\sum_k U_k W_k |\Psi_{ex}\> = PV |\Psi_{ex}\> = 0
\label{ukwk2}
\end{equation}
This is what we wanted to show.

{\bf Step 3:}
It follows from equations (\ref{ukwk1}) and (\ref{ukwk2}) that
\begin{align}
\sum_k \<\Psi| U_k W_k |\Psi\> = 1 \ , \ \sum_k \<\Psi_{ex} | U_k W_k |\Psi_{ex}\> = 0
\end{align}
In particular,         
\begin{align}
\sum_k \<\Psi| U_k W_k |\Psi\> \neq \sum_k \<\Psi_{ex}| U_k W_k | \Psi_{ex}\>
\label{uw}
\end{align}

At the same time, it is easy to see that
\begin{equation}
\<\Psi_{ex}|U_k W_k|\Psi_{ex}\> = \<\Psi |U_k W_k | \Psi \> \cdot e^{i\theta_{kl}}         
\label{braid}
\end{equation}
where $\theta_{kl}$ is the mutual statistics between $k$ and $l$. One way to derive this relation is
to use the representation $|\Psi_{ex}\> = W_{l\gamma} |\Psi\>$ (\ref{psiexeq}), and to choose the path $\gamma$ so
that it intersects the path corresponding to $W_k$ at one point (Fig. \ref{wlgammafig}b). Equation (\ref{braid}) then follows
immediately from the string commutation algebra (\ref{stringcomm}).

To complete the derivation, we now compare the two relations (\ref{uw}), (\ref{braid}). From (\ref{braid}),
we see that $\<\Psi| U_k W_k |\Psi\> = \<\Psi_{ex} |U_k W_k | \Psi_{ex}\>$ if $k$ and $l$ have trivial mutual
statistics. Also, $\<\Psi| U_k W_k |\Psi\> = \<\Psi_{ex} |U_k W_k | \Psi_{ex}\>$ if $\<\Psi|U_k W_k|\Psi\> = 0$. 
On the other hand, from (\ref{uw}), we know that $\<\Psi| U_k W_k |\Psi\> \neq \<\Psi_{ex} |U_k W_k | \Psi_{ex}\>$
for at least one $k$. We conclude that there must be at least one particle type $k$ which
has nontrivial statistics with respect to $l$ \emph{and} has $\<\Psi|U_k W_k |\Psi\> \neq 0$. Applying Lemma 1         
(\ref{lemmaeq}), we conclude that there is at least one particle $k$ which has nontrivial statistics with
respect to $l$ and can be annihilated at the boundary. Hence, the set $\mathcal{M}$ of quasiparticles that can
be annihilated at the boundary must obey condition (2) of the criterion, as claimed.

\section{Deriving the partition function transformation law} \label{transapp}
In this section we derive the transformation law (\ref{zltrans}) for $Z_l(\tau)$ using the Poisson summation formula. This 
computation is well-known (see e.g. Ref. \onlinecite{Francesco}), but we include it here for completeness. 

The first step is to derive an explicit expression for $Z_l(\tau)$. To this end, consider a general operator of the form
\begin{equation}
e^{i(l+K \Lambda)^T \phi} \mathcal{O}_{\{n_{J,k}\}} 
\end{equation}
Rewriting this operator in terms of the $\t{\phi}$ fields gives
\begin{equation}
e^{i \gamma^T \t{\phi}} O_{\{n_{J,k}\}}
\end{equation}
where $\gamma = W^T (l+ K \Lambda)$. The scaling dimensions for this operator are therefore
\begin{eqnarray}
\Delta(\gamma,\{n_{J,k}\}) &=&  \frac{1}{2} \gamma^T \bpm \bf{1} & 0 \\ 0 & 0 \epm \gamma + 
\sum_{J=1}^{N} \sum_{k=1}^\infty k \cdot n_{J,k} \label{scaldimgen} \\
\wb{\Delta}(\gamma, \{n_{J,k}\}) &=&  \frac{1}{2} \gamma^T \bpm 0 & 0 \\ 0 & \bf{1} \epm \gamma + 
\sum_{J=N+1}^{2N} \sum_{k=1}^\infty k \cdot n_{J,k} \nonumber
\end{eqnarray}
where $\bf{1}$ is an $N \times N$ identity matrix.

To calculate the partition function $Z_l(\tau)$, we substitute these scaling dimensions into
(\ref{partfunct}) and sum over all of the above operators. That is, we sum over all $\{n_{J,k}\}$ 
and all $\gamma \in \Gamma_l$ where $\Gamma_l$ denotes the lattice
\begin{equation}
\Gamma_l = \{W^T (l + K \Lambda) : \Lambda \in \mathbb{Z}^{2N} \}
\end{equation}
Simplifying the resulting sum, we find:
\begin{equation}
Z_l(\tau) = \sum_{\gamma \in \Gamma_l} e^{-\pi \gamma^T A(\tau) \gamma} \frac{1}{|\eta(\tau)|^{2N}}
\label{zlexpr}
\end{equation}
where $A(\tau)$ is the $2N \times 2N$ matrix
\begin{equation}
A(\tau) = \bpm -i \tau \cdot \bf{1} & 0 \\ 0 & i \wb{\tau} \cdot \bf{1} \epm
\end{equation}
and $\eta$ is the Dedekind eta function
\begin{equation}
\eta(\tau) = e^{\pi i \tau/12} \prod_{k=1}^\infty \left(1-e^{2\pi i k \tau} \right)             
\end{equation}  

We are now ready to derive the transformation law (\ref{zltrans}). First we use the Poisson summation formula to deduce
\begin{eqnarray}
\sum_{\gamma \in \Gamma_l} e^{-\pi \gamma^T A(-1/\tau) \gamma} 
&=& \frac{1}{\text{vol}(\Gamma_0)} \frac{1}{\sqrt{\det{A(-1/\tau)}}} \label{poiss} \\
\cdot \sum_{\gamma^* \in \Gamma_0^*}
&e&^{-\pi(\gamma^*)^T A(-1/\tau)^{-1} \gamma^* + 2\pi i l^T W \gamma^*}  \nonumber
\end{eqnarray}
Here $\text{vol}(\Gamma_0)$ denotes the volume of the unit cell of $\Gamma_0$ and $\Gamma_0^*$ denotes
the set of all vectors that have integer inner product with the vectors in $\Gamma_0$ (i.e. the dual lattice). 

Next, we observe that
\begin{align*}
A\left(-\frac{1}{\tau}\right)^{-1} = A(\tau) \ , \ \det{A\left(-\frac{1}{\tau}\right)} = \frac{1}{|\tau|^{2N}}  
\end{align*}
and $\text{vol}(\Gamma_0) = \sqrt{\det{K}}$. Also, it is not hard to show that $\Gamma_0^*$ can be written as 
\begin{equation}
\Gamma_0^* = \bigcup_{l' \in \mathcal{L}} \Sigma_z \cdot \Gamma_{l'}
\end{equation}
where $\Sigma_z = \bpm \bf{1} & 0 \\ 0 & -\bf{1} \epm$.
Substituting these expressions into equation (\ref{poiss}) and simplifying, we derive the transformation law
\begin{displaymath}
\sum_{\gamma \in \Gamma_l} e^{-\pi \gamma^T A (-1/\tau) \gamma} =
|\tau|^{N} \sum_{l' \in \mathcal{L}} S_{ll'} \sum_{\gamma \in \Gamma_{l'}} e^{-\pi \gamma^T A(\tau) \gamma}
\end{displaymath}
where
\begin{equation}
S_{ll'} = \frac{1}{\sqrt{\det{K}}} e^{i \theta_{ll'}}
\end{equation}

Combining this result with the identity
\begin{equation}
\eta(-1/\tau) = \sqrt{-i \tau} \cdot \eta(\tau) 
\end{equation}
we conclude that
\begin{equation}
Z_l(-1/\tau) = \sum_{l'} S_{ll'} Z_{l'}(\tau)
\end{equation} 
as claimed.

\section{Bosonic case} \label{boseapp}
As discussed in the conclusion, the conditions for when an abelian bosonic system has a gapped edge are similar to that of fermionic systems. 
The only difference is that the set of quasiparticles $\mathcal{M}$ must have an additional property beyond the two from the
fermionic case. Specifically, we require that every $m \in \mathcal{M}$ must be a boson, i.e. $e^{i \theta_m} = 1$. The bosonic
criterion states that a gapped edge is possible if and only if there exists a set $\mathcal{M}$ with these three properties (i.e. 
a ``Lagrangian subgroup''). In this section, we discuss how to modify the different arguments in this paper to prove
this claim.

\subsection{Microscopic argument}
As in the fermionic case, the starting point for the microscopic argument is the Chern-Simons theory (\ref{cstheory}) where 
$K_{IJ}$ is a $2N \times 2N$ symmetric, non-degenerate integer matrix. However, we now restrict ourselves 
to matrices $K_{IJ}$ whose diagonal elements are all \emph{even}, since we are interested in systems built out of bosons.

Following the same analysis as in section \ref{gensect}, it suffices to prove the following mathematical result:
there exists $N$ linearly independent integer vectors $\{\Lambda_1,...,\Lambda_N\}$ satisfying $\Lambda_i^T K \Lambda_j = 0$ 
if and only if there exists a set of (inequivalent) integer vectors $\mathcal{M}$ satisfying the following properties: 
\begin{enumerate}
\item{$m^T K^{-1} m'$ is an integer for any $m, m' \in \mathcal{M}$.}
\item{$m^T K^{-1} m$ is an even integer for any $m \in \mathcal{M}$.}
\item{If $l$ is not equivalent to any element of $\mathcal{M}$, then $m^T K^{-1} l$ is non-integer for some $m \in \mathcal{M}$.}
\end{enumerate}
Here the second property comes from the requirement that $e^{i\theta_m} = 1$ for every $m \in \mathcal{M}$.

The ``only if'' direction can be established exactly as in the fermionic case (appendix \ref{proofapp}) with no modification. 
The ``if'' direction is also quite similar to the fermionic case: as in appendix \ref{proofapp}, we define a $2N$ dimensional 
lattice $\Gamma$ by equation (\ref{gammadef}) 
and we construct a matrix $U$ with $\Gamma = U \mathbb{Z}^{2N}$. We then define $P = U^T K^{-1} U$. Just as before, it is easy to 
see that $P$ is a symmetric integer matrix with vanishing signature and unit determinant. The only difference from the 
fermionic case is that $P$ is now an ``even'' matrix instead of an ``odd'' matrix. Thus, Milnor's theorem implies that we can 
block diagonalize $P$ as
\begin{equation}
W^T P W = \bpm 0 & \bf{1} \\ \bf{1} & 0 \epm
\label{Peq2}
\end{equation}
instead of (\ref{Peq}). We then define $\Lambda_i = \det K \cdot K^{-1} U w_i $ where $w_i$ is the $i$th column of $W$, $i=1,...,N$. 
As in the fermionic case, it is easy to see that $\Lambda_i$ obey $\Lambda_i^T K \Lambda_j = 0$, and are linearly independent and integer.

It is also possible to prove the stronger correspondence of appendix \ref{strongapp} in the bosonic case. That is, it 
is possible to construct a gapped edge for each $\mathcal{M}$ in such a way that the quasiparticles in $\mathcal{M}$ can be 
annihilated at the boundary. The only difference from the fermionic construction of appendix \ref{strongapp}, is that in the bosonic case we 
need to consider an edge theory described by
\begin{equation}
K' = \bpm K & 0 & 0 \\
	  0 & 0 & -\mathbf{1} \\
	  0 &  -\mathbf{1} & 0  \epm
\end{equation}
instead of (\ref{kprime}). The rest of the analysis proceeds as before.

\subsection{Braiding statistics argument}
It is straightforward to extend the braiding statistics argument of section \ref{braidargsect} to the bosonic case. 
As before, we assume a gapped edge, and we define $\mathcal{M}$ to be the set of particles that can be annihilated at the boundary. 
Using arguments identical to the fermionic case, we can show that $\mathcal{M}$ satisfies the two properties discussed in
the introduction. The only new element is that we now have to show that $\mathcal{M}$ satisfies the additional property that 
every $m \in \mathcal{M}$ is a boson. Similarly to section \ref{braidargsect}, this statement can be established by constructing
string-like operators $\mathbb{W}_{m \beta} = U_a U_b W_{m \beta}$ and then examining their commutation rules. However, instead of 
using the commutation algebra (\ref{stringcomm}), one needs to use the hopping operator algebra of 
Ref. \onlinecite{LevinWenHop}:
\begin{equation}
W_{m \beta} W_{m \gamma} W_{m \delta} = e^{i\theta_m} W_{m \delta} W_{m \gamma} W_{m \beta}
\label{hopalg}
\end{equation}
Here $e^{i \theta_m}$ denotes the exchange statistics of $m$ and $\beta, \gamma, \delta$ are three open paths that share a common endpoint. Following an approach similar to section \ref{braidargsect},
one can show that self-consistency requires that $e^{i \theta_m} = 1$ for all $m \in \mathcal{M}$, as claimed.

\subsection{Modular invariance argument}
As in the fermionic case, the modular invariance argument begins by considering a strip geometry in which the lower edge is gapped 
while the upper edge described by the conformal field theory (\ref{cftedge}). We then define a partition function $Z(\tau)$ (\ref{partfunct}) 
and we show that $Z(\tau)$ can be written as a sum (\ref{partfunct2})
\begin{equation}
Z(\tau) = \sum_{m \in \mathcal{M}} Z_m(\tau) \label{partfunct3}
\end{equation}
where $\mathcal{M}$ denotes the set of quasiparticles that can be annihilated at the lower edge. Proceeding as in section 
\ref{modinvsect}, we can use the modular invariance constraint (\ref{modinv}) to show that $\mathcal{M}$
satisfies the two properties discussed in the introduction. The only new element is that we now have to show that $\mathcal{M}$ satisfies
the additional property that every $m \in \mathcal{M}$ is a boson. 

To establish this additional property, we make use of the second modular invariance constraint\cite{Francesco}
\begin{equation}
Z(\tau+1) = Z(\tau)
\label{modinvt}
\end{equation}
We then use the transformation law
\begin{align}
Z_l(\tau+1) = \sum_{l' \in \mathcal{L}} T_{ll'} Z_{l'}(\tau) \ \ ; \ \ T_{ll'} = e^{i \theta_l} \delta_{ll'}
\label{ttrans}
\end{align}
Like (\ref{zltrans}), this relation can be derived either from the explicit expression for $Z_l(\tau)$ (\ref{zlexpr}) or from the
general equivalence between the ``modular $T$-matrix''\cite{Francesco} and ``topological $T$-matrix.''\cite{Kitaevhoneycomb}
Substituting (\ref{ttrans}) into (\ref{partfunct3}), we can see that modular invariance 
requires that $e^{i\theta_m} = 1$ for all $m \in \mathcal{M}$, as claimed.

\bibliography{protedge}

\end{document}